\documentclass{article}
\usepackage[utf8]{inputenc}
\usepackage{graphicx}
\usepackage{caption}
\usepackage[left=2.5cm, right=2.5cm, bottom=2.5cm, top=2.5cm]{geometry}
\usepackage[square, numbers, comma, sort&compress]{natbib}
\usepackage{url}
\usepackage{amsmath}
\usepackage{svg}
\usepackage{booktabs}
\usepackage{siunitx}
\usepackage{hyperref}
\usepackage{comment}
\usepackage{subcaption}
\usepackage{multirow}
\usepackage{authblk}
\usepackage{xcolor}

\title{Impact of phylogeny on structural contact inference from protein sequence data}
\author{Nicola Dietler\textsuperscript{1,2}, Umberto Lupo\textsuperscript{1,2}, Anne-Florence Bitbol\textsuperscript{1,2,*}}
\affil{\textbf{1} Institute of Bioengineering, School of Life Sciences, École Polytechnique Fédérale de Lausanne (EPFL), CH-1015 Lausanne, Switzerland\\
\textbf{2} SIB Swiss Institute of Bioinformatics, CH-1015 Lausanne, Switzerland\\
* Corresponding author:  \href{mailto:anne-florence.bitbol@epfl.ch}{anne-florence.bitbol@epfl.ch}}
\date{}

\begin{document}

\maketitle
%The abstract should be no more than 200 words , 8000 words for article.
%Please include at least 3 and up to 6 keywords.

\begin{abstract}
Local and global inference methods have been developed to infer structural contacts from multiple sequence alignments of homologous proteins. They rely on correlations in amino-acid usage at contacting sites. Because homologous proteins share a common ancestry, their sequences also feature phylogenetic correlations, which can impair contact inference. We investigate this effect by generating controlled synthetic data from a minimal model where the importance of contacts and of phylogeny can be tuned. We demonstrate that global inference methods, specifically Potts models, are more resilient to phylogenetic correlations than local methods, based on covariance or mutual information. This holds whether or not phylogenetic corrections are used, and may explain the success of global methods. We analyse the roles of selection strength and of phylogenetic relatedness. We show that sites that mutate early in the phylogeny yield false positive contacts. We consider natural data and realistic synthetic data, and our findings generalise to these cases. Our results highlight the impact of phylogeny on contact prediction from protein sequences and illustrate the interplay between the rich structure of biological data and inference. 
\end{abstract}

%Statistical inference from biological data is currently making great advances, thanks to the rapid growth of available data, and to the development and application of inference methods. It is important to understand the impact of the rich and complex structure of biological data on the performance of inference methods. We study how correlations coming from shared ancestry impact the inference of structural contacts from multiple sequence alignments of homologous proteins.  We find that inferred Potts models are more resilient to phylogenetic correlations than covariance or mutual information. This contributes to the success of these models on natural protein sequences.

\section*{Introduction}

Statistical inference from biological data is currently making great advances, thanks to the rapid growth of available data, and to the development and application of inference methods. These methods range from interpretable models inspired by statistical physics to deep learning approaches. In this exciting context, it is important to understand the impact of the rich and complex structure of biological data on the performance of inference methods~\cite{Ngampruetikorn22}. 

In particular, progress in sequencing has caused a spectacular growth of available genome sequences. This allows inference to be performed on large multiple sequence alignments (MSAs) of families of homologous proteins. Because proteins in the same family share the same ancestry, as well as similar three-dimensional structures and functional properties, this MSA data is highly structured. In particular, MSAs comprise correlations coming from phylogeny~\cite{Casari95,Halabi09,Qin18}, as well as from structural and from functional constraints~\cite{Gobel94,Pazos97}. Pairwise maximum entropy models~\cite{Jaynes57,Lapedes99}, known as Potts models or Direct Coupling Analysis (DCA)~\cite{Weigt09}, trained on natural MSAs, have revealed structural contacts~\cite{Marks11,Morcos11,Sulkowska12}. They have also been employed to analyze mutational effects~\cite{Dwyer13,Cheng14,Cheng16,Figliuzzi16}, protein evolution~\cite{delaPaz20} and conformational changes~\cite{Morcos13,Malinverni15}, to design proteins~\cite{Russ20}, as well as to predict interaction partners among paralogs~\cite{Bitbol16,Gueudre16} and protein-protein interaction networks~\cite{Cong19,Green21}. The key idea of these fruitful approaches is coevolution: amino acids that are in contact in the three-dimensional structure of proteins need to maintain physico-chemical complementarity, yielding correlations in amino-acid usage at contacting sites. Recently, AlphaFold, a deep learning approach exploiting the breadth of available sequences and experimentally-determined structures, has brought major advances to the computational prediction of protein structures from sequences~\cite{Jumper21}. Despite the important methodological differences with DCA, coevolution between amino acids is also an important ingredient of Alphafold, which starts by constructing an MSA of homologs when given a protein sequence as input~\cite{Jumper21,RoneyPreprint}.

How does the complex structure of MSA data impact the performance of statistical inference?
In MSAs, correlations due to phylogeny~\cite{Casari95,Qin18} and due to selection to preserve structure~\cite{Marks11,Morcos11,Sulkowska12} and function coexist. Thus, phylogenetic correlations can impair the inference of structural contacts. This effect has been demonstrated both when using mutual information~\cite{Dunn08} and when using inferred Potts models~\cite{Qin18,Vorberg18,RodriguezHorta19,RodriguezHorta21}. It has motivated the development and use of various empirical corrections aimed at reducing the impact of phylogenetic correlations~\cite{Lichtarge96,Dunn08}, such as the Average Product Correction~\cite{Ekeberg13,Hockenberry19}, phylogenetic reweighting~\cite{Weigt09,Marks11,Morcos11,Malinverni20,Hockenberry19}, and Nested Coevolution~\cite{Colavin22}. 

How do phylogenetic correlations impact MSA properties and impair contact prediction using mutual information, covariance or inferred Potts models?  
%Minimal models have been used to address the assessment of the Potts model performance on structural prediction or interaction partners prediction without phylogeny in~\cite{Ngampruetikorn22,Gandarilla20}. The impact of phylogeny has also been studied on DCA performance with a different approach in~\cite{RodriguezHorta21} and with a similar approach but on prediction of interacting partners in~\cite{Gerardos22}. Therefore, we ask the question of 
To address this question, we generate synthetic data from a minimal model where the amount of correlations from structural contacts and from phylogeny can be fully controlled. We analyse the impact of the strength of natural selection to preserve structure, the impact of phylogenetic relatedness, and their interplay. While phylogeny impairs contact prediction, we show that Potts models are more robust to phylogeny than local scores, even in the absence of explicit phylogenetic corrections, which are useful in both cases. This robustness to phylogeny comes on top of the ability of Potts models to disentangle direct and indirect correlations (which only matters for strong selection), and may explain the success of these methods. We further show that sites that mutate early in the phylogeny yield false positive contacts. Next, we analyse natural protein sequence data and realistic synthetic data generated from models inferred on natural data, either with or without phylogeny. Our findings from the minimal model generalise well to these cases.

\section*{Results}

\begin{figure}[htb]
\centering
\includegraphics[width=0.9\textwidth]{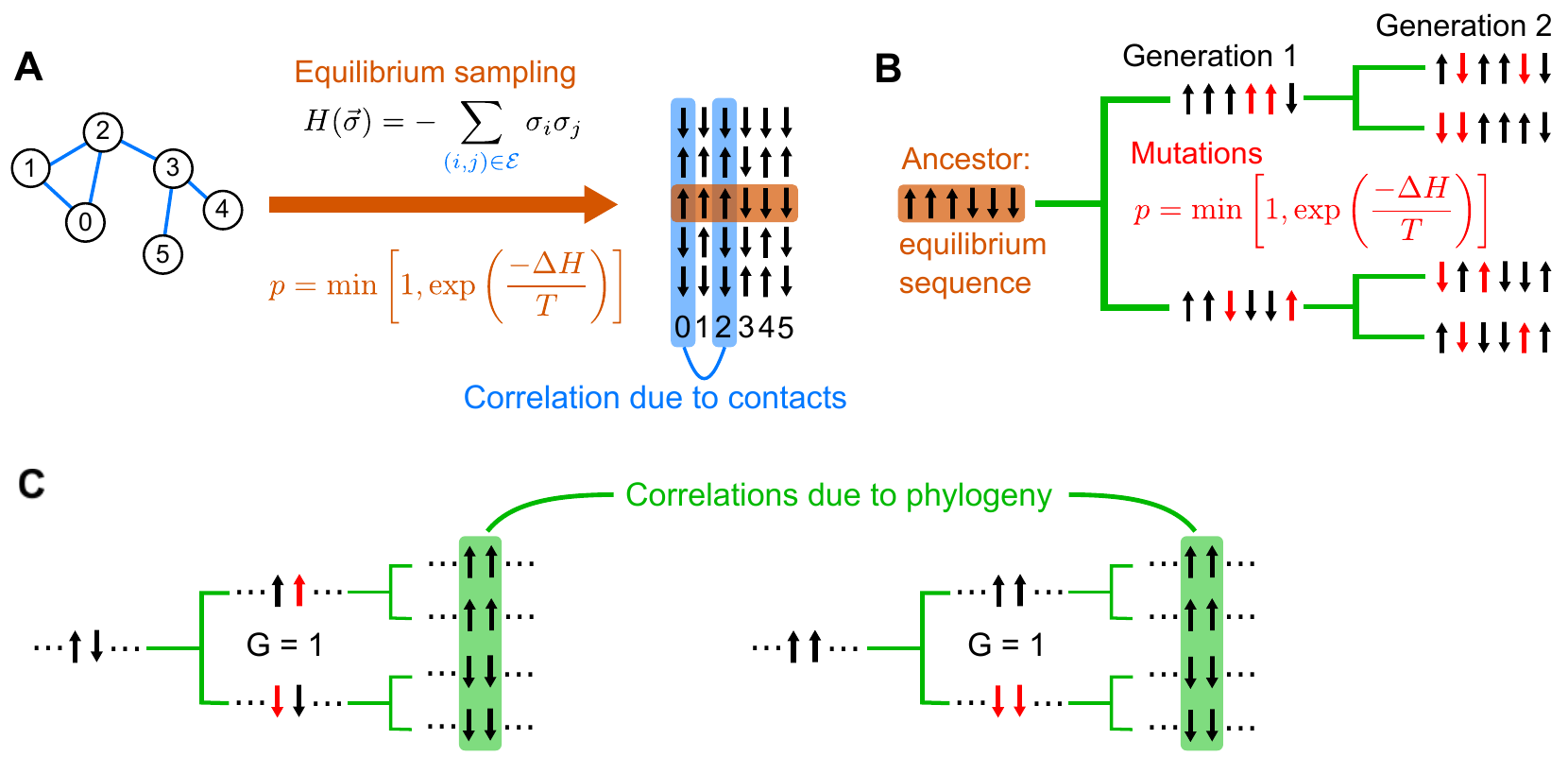}
\caption{\textbf{Data generation and origins of correlations}. \textbf{A:} In our model, couplings between nodes are set to 1 on the edges of an Erd\H{o}s-Rényi random graph, and to 0 otherwise (the set of edges is denoted by $\cal E$). Placing a spin $\sigma_i$ on each node of the graph gives rise to the Hamiltonian $H$ for a sequence $\vec{\sigma}$ of spins. Equilibrium sampling of sequences (without phylogeny) is performed using a Metropolis Monte Carlo algorithm, with move acceptance probability $p$ defined as shown, starting from random initial sequences. An example correlation arising from a coupling is shown between columns $0$ and $2$ in the resulting set of independently generated equilibrium sequences (blue). \textbf{B:}  To generate sequences with phylogeny, an equilibrium sequence (generated as in \textbf{A}) is taken as the ancestor, and is evolved along a binary branching tree (green) where mutations (red) are accepted with probability $p$. On each branch of the tree, $\mu$ mutations are accepted, with $\mu=2$ here. \textbf{C:}  Examples of phylogenetic correlations. On the left hand side, early mutations arise at two sites in different sequences, leading to highly correlated sites in their daughter sequences (assuming no further mutations at those sites). On the right hand side, early mutations arise simultaneously in one sequence, also giving a large correlation between theses sites in the daughter sequences. In both cases, the earliest mutation generation $G$ of the pair of sites of interest is given. It is defined as the first generation in the tree at which both sites have mutated with respect to their state in the ancestral sequence.}
\label{fig:figure_conceptual}       
\end{figure}

\paragraph{Minimal model with structural constraints and phylogeny.} Separating correlations from structural contacts and from phylogeny is extremely tricky in natural data~\cite{WeinsteinPreprint}. Thus, to assess the impact of phylogeny on structural contact prediction, we generate and study synthetic data with controlled amounts of correlations from structural constraints and from phylogeny. For this, we consider a minimal model, where protein sequences are represented by sequences of Ising spins, each spin sitting on one node of a fixed Erd\H{o}s-Rényi random graph. Structural constraints, representing contacts in the three-dimensional protein structure, are modelled by pairwise couplings on the edges of this graph. We then sample independent equilibrium sequences using a Metropolis--Hastings algorithm (see Methods and Figure~\ref{fig:Magnetisation_vs_tau}). This yields data sets of sequences that are only subject to the structural constraints defined by the graph. These data sets are regarded as multiple sequence alignments (MSAs). All correlations between sites then arise from these constraints or from finite size effects. In order to introduce phylogeny in a controlled way, we start from one equilibrium sequence, considered as the ancestral sequence, and we evolve sequences on a binary branching tree with a fixed number $\mu$ of accepted mutations, i.e.\ substitutions, per branch. Mutations are accepted using the same Metropolis criterion as when generating independent equilibrium sequences. The sequences at the leaves of the tree yield an MSA that incorporates structural constraints and phylogeny. Both of them can yield correlations between the columns of the MSA. Our method to generate data is illustrated in Figure~\ref{fig:figure_conceptual}, and described in detail in Methods. Note that the case with phylogeny and no selection is a limiting case of our model (see below). 

\begin{figure}[htbp]
\centering
\includegraphics[scale = 0.4]{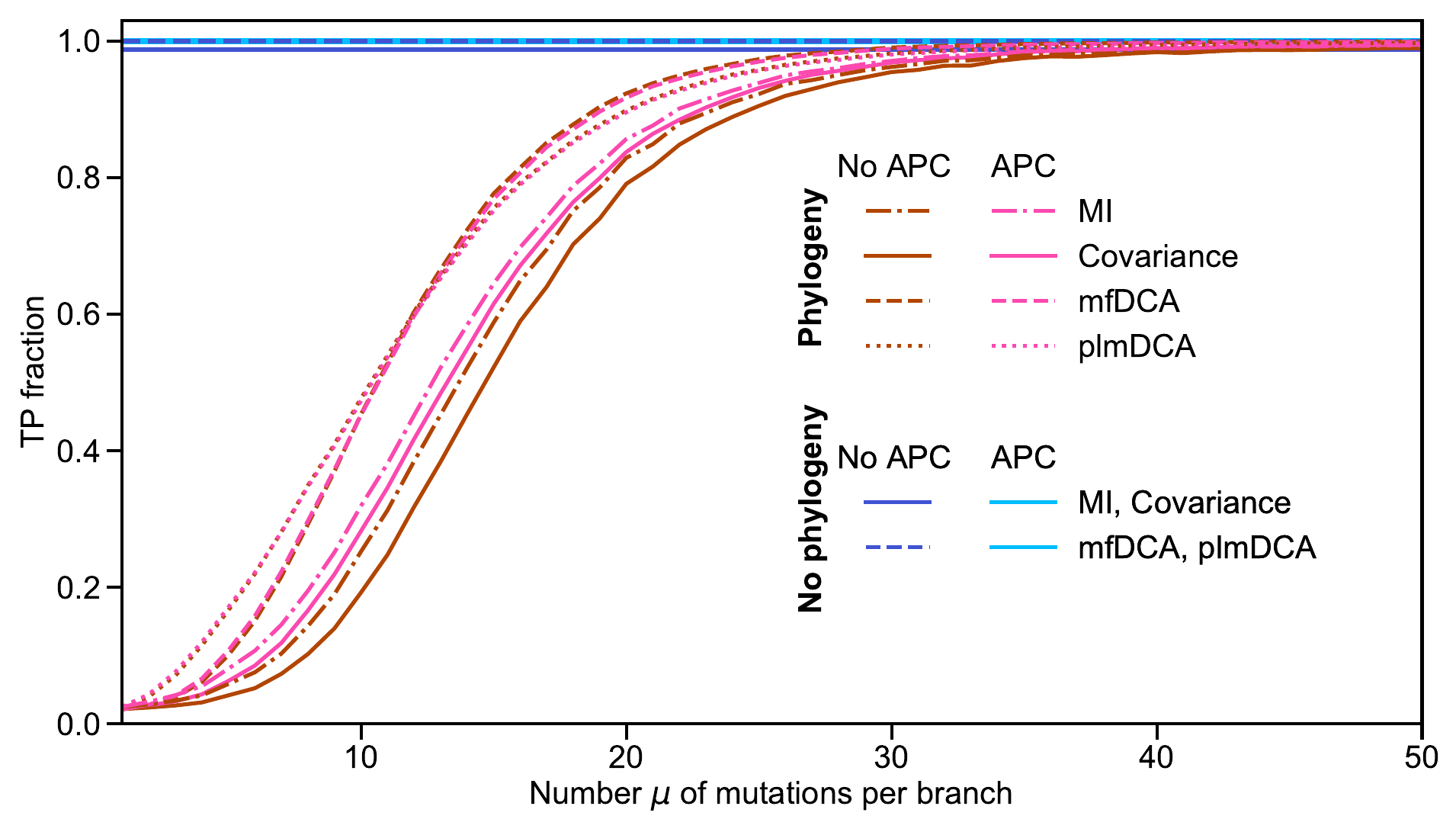}
\caption{\textbf{Impact of phylogeny on contact prediction.} The fraction of correctly predicted contacts (True Positive (TP) fraction) is shown versus the number $\mu$ of  accepted mutations (i.e.\ substitutions) per branch of the phylogenetic tree. Four different inference methods are compared: Covariance, Mutual Information (MI), mfDCA and plmDCA. In all cases, results are presented without and with Average Product Correction (APC).
The data generated without phylogeny, used for reference, comprises $M = 2048$ sequences of length $\ell = 200$ sampled at equilibrium by Metropolis Monte Carlo at $T=5$ under the Hamiltonian in Eq.~\ref{miniHam} on an Erd\H{o}s-Rényi graph with edge probability $q = 0.02$, representing the contact map. Data with phylogeny is generated starting from one of the equilibrium sequences, and this ``ancestor'' is evolved along a binary branching tree, with a fixed number $\mu$ of accepted mutations per branch, for $11$ generations, yielding $M=2^{11}=2048$ sequences. Random proposed mutations (spin flips) are accepted according to the Metropolis criterion at $T = 5$ under the Hamiltonian in Eq.~\ref{miniHam} on the same Erd\H{o}s-Rényi graph. All results are averaged over 100 realisations (i.e.\ 100 data generations, always for the same Erd\H{o}s-Rényi graph).  Note that the standard deviation of the TP fraction is always smaller than $10^{-2}$. Note also that inference with APC on the data generated without phylogeny is as good for local and global methods, so they are shown with the same linestyle and colour.}
\label{fig:tpfrac_vs_mu_sparse}
\end{figure}

\paragraph{Impact of phylogeny on contact prediction.} What is the impact of phylogeny on contact prediction in this minimal model? Varying $\mu$ allows us to tune the amount of phylogeny. A small $\mu$ means that sequences are very closely related, while independent equilibrium sequences are recovered in the limit $\mu\to\infty$. In Figure~\ref{fig:tpfrac_vs_mu_sparse}, we show the performance of contact prediction versus $\mu$. This performance is quantified by the True Positive (TP) fraction of predicted contacts, evaluated at the number of edges in the graph (see Methods), and we compare it to that obtained for equilibrium data sets. We perform contact inference with four different methods: two local ones, based on covariance and mutual information (MI) between sites, and two global ones, based on inferring DCA Potts models that fit the one- and two-body frequencies of the data. Local methods directly compare pairwise statistics, while global methods aim to infer a probability distribution for the whole sequence. The two global methods, mean-field DCA (mfDCA)~\cite{Marks11,Morcos11} and pseudo-likelihood maximization DCA (plmDCA)~\cite{Ekeberg13,Ekeberg14} (see Methods), employ different approximation schemes in the Potts model inference.
In all cases, we present results both with and without the Average Product Correction (APC), which was introduced to correct for biases due to conservation and phylogeny when using mutual information to predict structural contacts~\cite{Dunn08}. This correction improves contact inference performance by DCA on natural protein sequence data~\cite{Ekeberg13,Ekeberg14}, and is widely used. Furthermore, we find that it improves contact prediction performance substantially more than phylogenetic reweighting when using synthetic data generated from models inferred on natural data (see below). For equilibrium data, all inference methods yield a very good performance, with TP fractions close to $1$. For data generated with phylogeny, the equilibrium results are recovered for large $\mu$, as expected. However, inference performance is substantially impaired for smaller values of $\mu$ (corresponding to strong phylogeny). Importantly, global (DCA) methods are more resilient to phylogenetic noise than local ones (Covariance and MI) for small and intermediate values of $\mu$. This result is robust to whether APC is used or not, but we note that APC improves the performance of local methods when phylogeny is strong, while it has almost no effect on the performance of global methods.  For instance, for  $\mu$ = 15, the TP fraction obtained by plmDCA is 28\% (resp.\ 17\%) higher than that obtained by MI without (resp.\ with) APC. APC generally improves the performance of both local~\cite{Dunn08} and global methods~\cite{Ekeberg13,Ekeberg14} on natural sequence data (see also our results below on natural and more realistic data). The success of APC with global methods may then be partly due to its ability to correct for conservation effects~\cite{Dunn08}, corresponding to non-zero fields in Potts models. While, in our minimal model, there is a sharp separation between non-contacts and contacts, inferred DCA models comprise many small nonzero couplings, due to phylogeny, functional constraints and finite-size effects. Our result that DCA methods are more robust to phylogeny than local methods still holds when a background of nonzero couplings is included, see Figure~\ref{fig:tpfrac_vs_mu_gaussianJij}.

\begin{figure}[htbp]
\centering
\includegraphics[scale = 0.4]{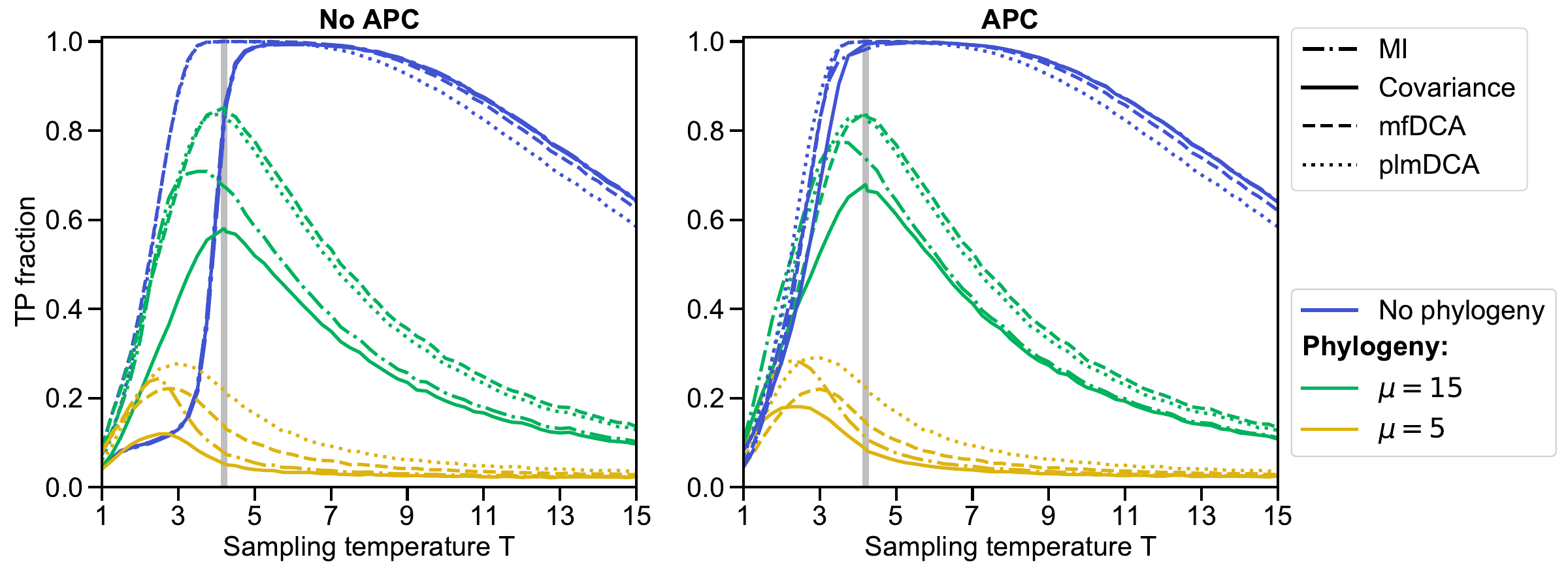}
\caption{\textbf{Impact of sampling temperature on contact prediction.} The TP fraction is shown versus the Monte Carlo sampling temperature $T$ for four different inference methods (Covariance, MI, mfDCA, plmDCA). The left panel shows results without APC, and the right panel with APC. Three cases are considered: equilibrium data (no phylogeny), and numbers of accepted mutations per branch $\mu=15$ and $\mu=5$ (increasing impact of phylogeny). Data is generated as in Figure \ref{fig:tpfrac_vs_mu_sparse}, using the same contact map (Erd\H{o}s-Rényi graph). The gray vertical line shows the approximate phase transition temperature $T_C$. All results are averaged over 100 realisations.}
\label{fig:tpfrac_vs_T_sparse}
\end{figure}

\paragraph{Impact of selection strength on contact prediction without phylogeny.} In addition to $\mu$, another important parameter of our model is the Monte Carlo sampling temperature $T$ (see Figure~\ref{fig:figure_conceptual} and Eq.~\ref{MetroCrit}). It plays the role of an inverse selection strength, where selection aims to preserve structure. Note that other selective pressures are not considered in our minimal model. For large $T$, almost all mutations are accepted, and structural constraints are weak. Therefore, in the limit $T\to\infty$, selection vanishes and all correlations are from phylogeny. Conversely, for small $T$, essentially only the mutations that decrease the energy of the sequence of spins are accepted, and structural constraints become stringent. Figure~\ref{fig:tpfrac_vs_T_sparse} shows the performance of contact prediction versus $T$ for equilibrium sequences (no phylogeny), and for sequences generated with $\mu = 15$ and $\mu = 5$. At equilibrium, our model possesses a ferromagnetic-paramagnetic phase transition, whose approximate temperature $T_C=4.2$ was found by inspecting magnetisation histograms~\cite{Gandarilla20}. For equilibrium sequences at low $T$, inference performance without APC is poor (Figure \ref{fig:tpfrac_vs_T_sparse}, left panel), and DCA methods perform substantially better than local methods (which are overlapping). The performance of DCA methods increases with $T$, and reaches a perfect score before $T_C$. Without APC, local methods also reach a perfect contact prediction score, but after $T_C$. However, they perform slightly better than DCA at higher temperatures. These trends are consistent with those previously described in Refs.~\cite{Gandarilla20,Ngampruetikorn22}, which analysed synthetic data generated independently at equilibrium, i.e.\ without phylogeny. Inference is difficult either if sequences are all frozen and redundant (very small $T$) or if sequences are too noisy (very large $T$), yielding better performance for intermediate values of $T$. Furthermore, APC substantially improves the performance of local methods at equilibrium, and makes their performance comparable to that of global methods even at small $T$ (Figure~\ref{fig:tpfrac_vs_T_sparse}, right panel). In the Supplementary Material, section~\ref{conservlowT}, we show that the weak performance of local methods on equilibrium data at low $T$ and without APC is associated to conservation properties (see Figure~\ref{fig:TP_vs_T_transfdata}), partly mitigated by regularisation in global methods. Furthermore, in section~\ref{comp}, we present a detailed comparison with Ref.~\cite{Ngampruetikorn22}, which shows that our results are fully consistent with theirs, and that regularisation is crucial to the performance of DCA (see Figure~\ref{fig:TPAUC_vs_T_densegraph_eqm15}).

\paragraph{Interplay between phylogeny and selection strength.}
What is the impact of phylogeny on the temperature dependence of contact prediction performance? As for equilibrium data, the best performance is obtained for intermediate values of $T$ with phylogenetic data, see Figure~\ref{fig:tpfrac_vs_T_sparse}. However, the maximum of performance shifts to lower values of $T$ when phylogeny is increased. This may be because, when phylogeny is strong, stringent selection (low $T$) keeps contacts relevant, thereby compensating for phylogeny, while at high $T$, the combination of thermal noise and phylogeny makes contact inference particularly tricky. Moreover, with phylogeny, DCA methods yield substantially better results than local methods for intermediate to high values of $T$, by contrast with the equilibrium case. This holds whether or not APC is used, even though APC improves the performance of local methods. Note however that MI slightly outperforms DCA methods at small values of $T$. Overall, the fact that DCA outperforms local methods in a wide range of selection strengths in the presence of phylogeny (generalising the result observed at $T=5$ in Figure \ref{fig:tpfrac_vs_mu_sparse}) is much better in line with observations on natural data~\cite{Morcos11} than with results obtained without phylogeny, where DCA yields at best marginal improvement compared to local methods with APC. Thus, robustness to phylogeny may be what makes DCA superior to local methods. We also note that plmDCA outperforms mfDCA when phylogeny is strongest ($\mu=5$), which is reminiscent of the results obtained on natural data~\cite{Ekeberg13}.   In the Supplementary Material, section~\ref{comp}, we consider a denser Erd\H{o}s-Rényi graph, and our results are robust to this change (see Figures~\ref{fig:TPAUC_vs_T_densegraph_eqm15} and~\ref{fig:TPAUC_vs_T_densegraph_m5m15}). However, larger values of pseudocounts or regularisation strength than the usual ones are then helpful (see also Figure~\ref{fig:tpfrac_vs_pseudocounts}). The contact density used throughout and this denser one are both in the range observed in natural proteins (see Table~\ref{tab:MSA}). 

\paragraph{Impact of graph properties on contact inference.} Before exploring in more detail the impact of phylogeny, let us analyse the impact of graph properties, reflecting structural constraints, on equilibrium results. Indeed, in phylogenetic data, this structural signal needs to be disentangled from phylogenetic correlations. For each pair of sites, we consider the shortest path length $L$ in the graph connecting the two sites -- $L = 1$ denotes contacting sites. Normalised histograms of coevolution scores for all pairs of sites are shown versus $L$ in Figure~\ref{fig:HistogramsSPL} without APC and in Figure~\ref{fig:HistogramsSPL_APC} with APC. For $T = 3<T_C$, without APC and especially for local methods, histograms of pairs in contact ($L=1$) have substantial overlap with other histograms, especially with those for $L = 2$, showing that indirect correlations impair inference. However, APC strongly mitigates this issue, allowing local methods to perform almost as well as global ones, as seen in Figure~\ref{fig:tpfrac_vs_T_sparse}. Close to $T_C$, histograms of coevolution scores feature no overlap for DCA methods, leading to perfect inference. The small overlaps that still exist for local methods are resolved by using APC. Thus, APC enables local methods to disentangle direct and indirect correlations at equilibrium almost as well as global ones. At $T = 5>T_C$, all methods yield perfect inference, as indirect correlations become weaker in the paramagnetic phase. Note also that plmDCA performs better than mfDCA at separating pairs with $L=1$ and $L=2$. 

In addition to $L$, we consider another descriptor of graph structure, namely the number $N$ of nearest neighbours of each site. To assess the impact of $N$ on the variance of equilibrium data, we perform principal component analysis (PCA) focusing on the matrix of covariances between sequences~\cite{Casari95}. Figure~\ref{fig:PCA_NN} shows the coordinates of sites on the first principal component versus $N$. At all temperatures considered, and most strongly at low ones, these coordinates are correlated with $N$, showing that connectivity is a crucial ingredient of the variance of sites. Indeed, groups of connected spins tend to be aligned together at low temperatures, while isolated spins are independent. This difference is strongest at low $T$ (see Eq.~\ref{MetroCrit}).

\begin{figure}[htbp]
\centering
\includegraphics[scale = 0.5]{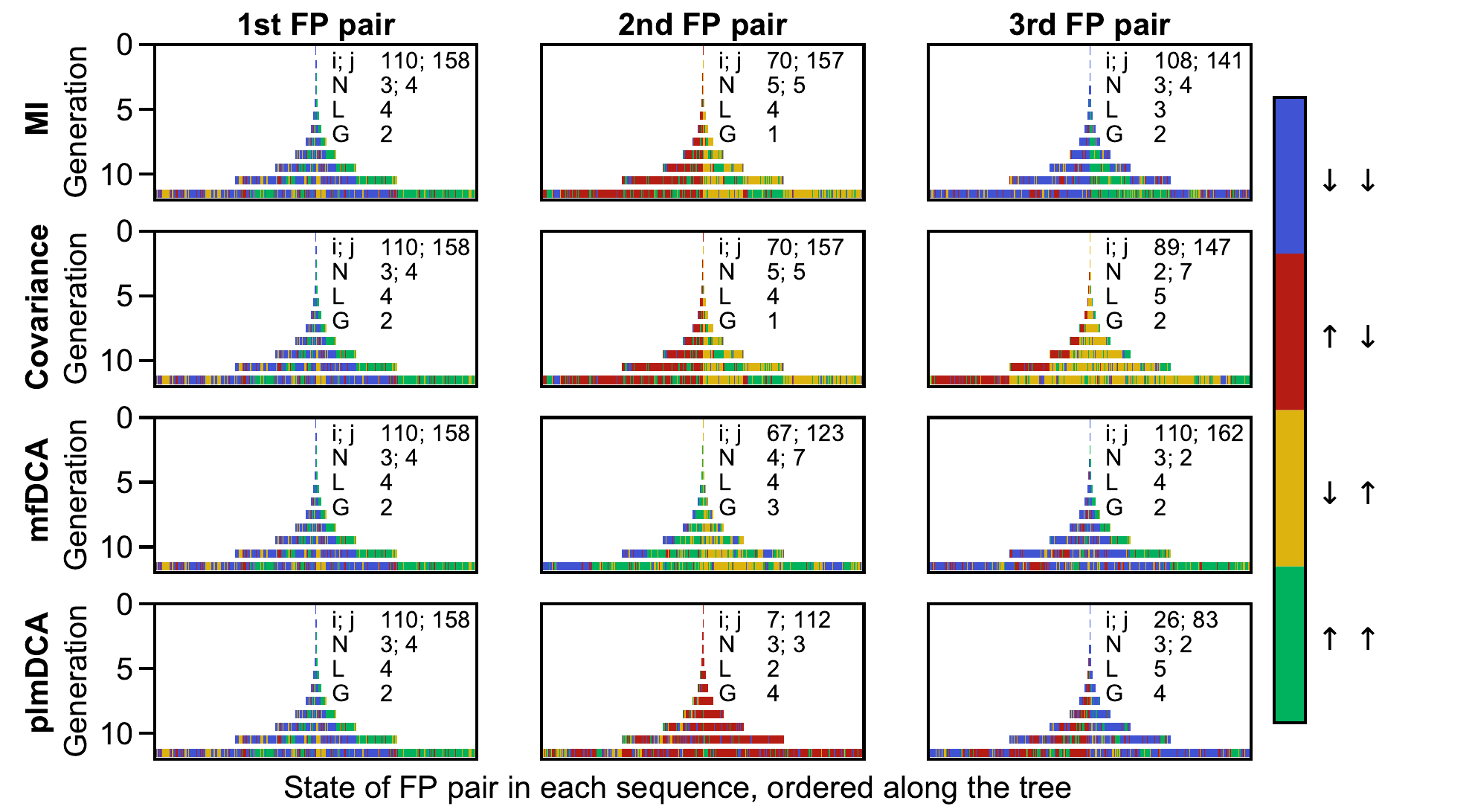}
\caption{\textbf{Evolution of false positive pairs along the phylogenetic tree.} For each inference method (MI, Covariance, mfDCA, plmDCA), the state of the top three false positive (FP) pairs of sites $(i,j)$, wrongly inferred as contacts from the evolved sequences, is depicted along the phylogenetic tree. This allows to track the history of these pairs $(i,j)$. At each generation, each vertical bar represents one sequence, and the colour of the bar indicates the state of the false positive pair of focus within this sequence. Hence, the ancestor sequence is the single bar at generation 0, and the final generation (11) contains 2048 bars corresponding to all evolved sequences. Data is generated as in Figure~\ref{fig:tpfrac_vs_mu_sparse}, using the same contact map (Erd\H{o}s-Rényi graph), at $T=5$ and $\mu=5$ (strong phylogeny). Characteristics of each FP pair $(i,j)$ are reported, namely the indices of the sites $i$ and $j$, their number $N$ of neighbours in the graph (i.e., sites in contact with $i$ or $j$), the length $L$ of the shortest path connecting $i$ to $j$ in the graph ($L=1$ for contacts), and the earliest mutation generation $G$ where both $i$ and $j$ have mutated with respect to the ancestral sequence.}
\label{fig:EvolutionPairsTree}
\end{figure}

\paragraph{Origin and impact of phylogenetic correlations.}
How does phylogeny impair contact inference? What pairs of sites appear to be coupled due to phylogeny, while they are actually not coupled? To address these questions, we focus on the false positive (FP) pairs which have the highest scores. Our synthetic data makes it possible to investigate the states of these pairs of sites through the phylogenetic tree, starting from the ancestral sequence. The states of the top three FP pairs found by each method are shown in Figure~\ref{fig:EvolutionPairsTree} for a data set generated at $\mu =5$ and $T=5$ (i.e.\ with strong phylogeny). We observe large blocks of colours in our representation, meaning that these pairs of sites underwent mutations early in the phylogeny, leading to strongly correlated sites in the last generation. Indeed, early mutations at two sites produce correlations among them, as illustrated in Figure~\ref{fig:figure_conceptual}B. This effect can be quantified by attributing to each pair of sites an index $G$, which is the first generation at which each of the spins of this pair has flipped (i.e.\ mutated), in at least one of the sequences generated thus far, with respect to its state in the ancestral sequence (see Figure \ref{fig:figure_conceptual}B). All top FP pairs in Figure~\ref{fig:EvolutionPairsTree} have small $G$ values, and this effect is even stronger for local methods ($G=1$ or $2$) than for DCA methods ($G=2$ to $4$). Thus, pairs of sites mutating early in the phylogeny are more likely to have high coevolution scores due to phylogeny. Their shortest path length $L$ is also indicative on the phylogenetic origin of their high coevolution scores. Indeed, $L > 2$ for all of them but one, while if indirect correlations were the main cause of these FPs, we would expect more occurrences of $L=2$. Note that, in Ref.~\cite{Gerardos22}, we showed that these pairs of sites are particularly useful for the inference of interacting partners among the paralogs of two protein families, where phylogenetic correlations are useful.

\begin{figure}[htbp]
\centering
\includegraphics[scale = 0.4]{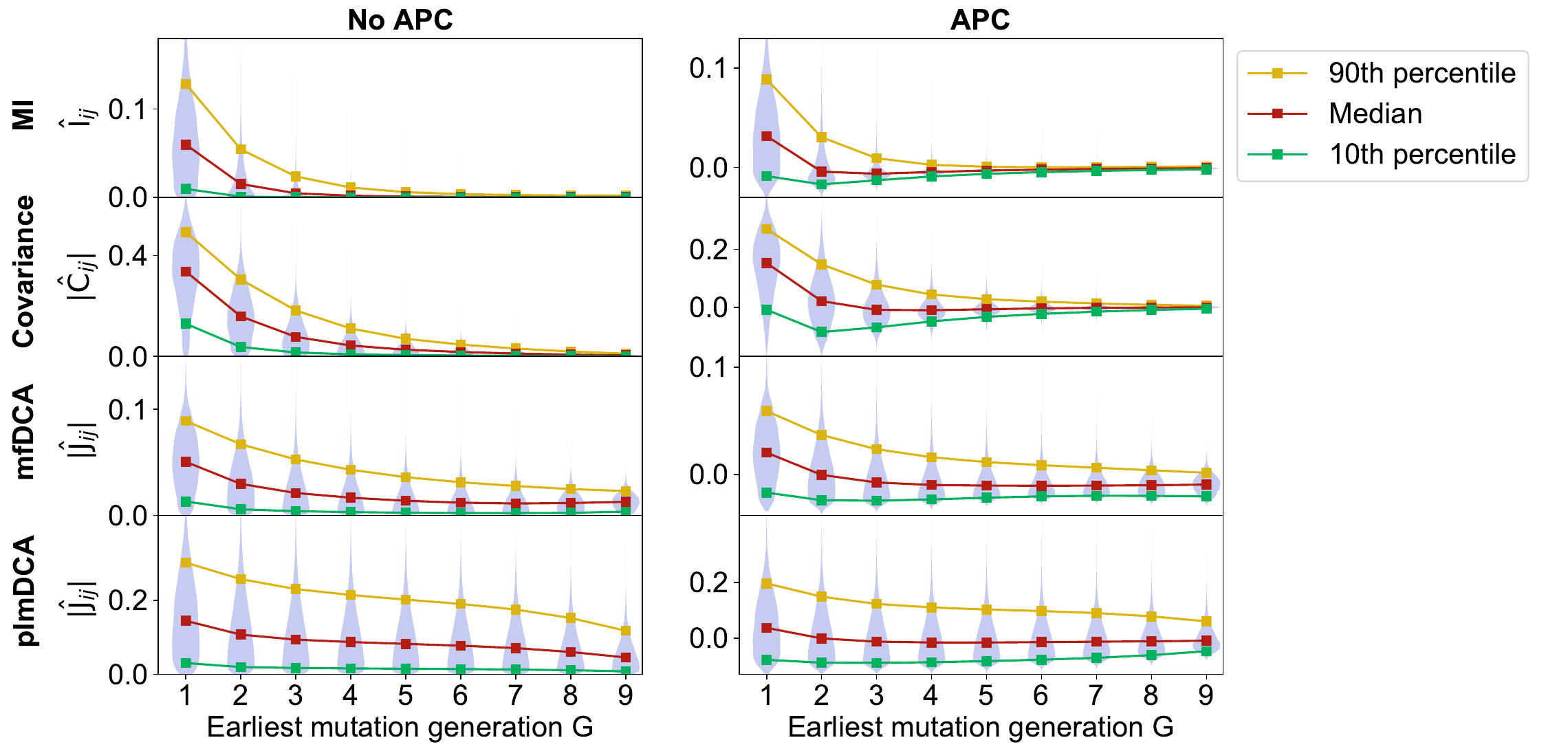}
\caption{\textbf{Impact of phylogeny on coevolution scores.} Violin plots of the coevolution scores of all pairs of sites $(i,j)$, inferred by four different methods (MI, Covariance, mfDCA, plmDCA), are shown versus the earliest generation $G$ where both $i$ and $j$ have mutated with respect to the ancestral sequence. The left panels show the raw Frobenius norm scores (absolute value here) and the right panels show these scores corrected with APC. Data is generated as in Figure~\ref{fig:tpfrac_vs_mu_sparse}, using the same contact map (Erd\H{o}s-Rényi graph), at $T=5$ and $\mu=5$. Couplings were inferred on 100 data sets of 2048 sequences each, and aggregated.}
\label{fig:Allcouplings_vs_emg}
\end{figure}

How do these observations made on a few top FPs generalise? To assess this, we compute the index $G$ of all pairs of sites $(i,j)$ -- including both contacts and non-contacts -- in multiple synthetic data sets generated at $\mu =5$ and $T=5$. (Note that the results that follow also hold when restricting to all FPs or to all non-contacts, which are much more numerous than contacts.) We then group pairs of sites that have a given value of $G$. Figure~\ref{fig:Allcouplings_vs_emg} shows the distribution of coevolution scores versus the index $G$ as a violin plot (i.e.\ Gaussian kernel-smoothed vertical histograms). It sheds light on an overall dependence of the coevolution scores on $G$, with larger coevolution scores observed for small $G$, i.e.\ early mutations. Moreover, this dependence is weaker for global methods than for local ones, and among global methods, it is weaker for plmDCA than for mfDCA. These conclusions hold both without and with APC. This confirms that DCA, especially plmDCA, is more robust to phylogenetic correlations than local inference methods, which could partly explain its success on natural protein sequences. The behaviour of the median score versus $G$ (red curve in Figure~\ref{fig:Allcouplings_vs_emg}) is further investigated in Figure~\ref{fig:median_vs_emg_fit}. Without APC, the median decays abruptly with $G$, especially for local methods. This decrease is less abrupt and closer to linear for plmDCA, confirming its lower sensitivity to the phylogeny. APC substantially reduces the median for small $G$ for all methods, thereby mitigating the impact of phylogeny on inference performance. 

Besides, Figure~\ref{fig:Averaged_conservation_vs_emg} shows the average conservation of sites involved in a pair versus $G$. It shows that pairs of sites undergoing mutations early in the phylogeny are often weakly conserved. Qualitatively, sites with early mutations will often be in different states (up and down) in sequences at the leaves of the tree, yielding a poor conservation but a strong correlation. This point could be exploited to disentangle TP and FP pairs when the underlying phylogeny is not known. The inferred fields could also be useful to this end. However, we expect the general case to be more complex as conservation can arise both from single-site constraints modelled by nonzero fields and from late divergences along the phylogeny. Note that in sector analysis, the focus is on conserved correlations, thus exploiting both aspects~\cite{Halabi09,Rivoire16}.

\begin{figure}[htb]
\centering
\includegraphics[scale = 0.4]{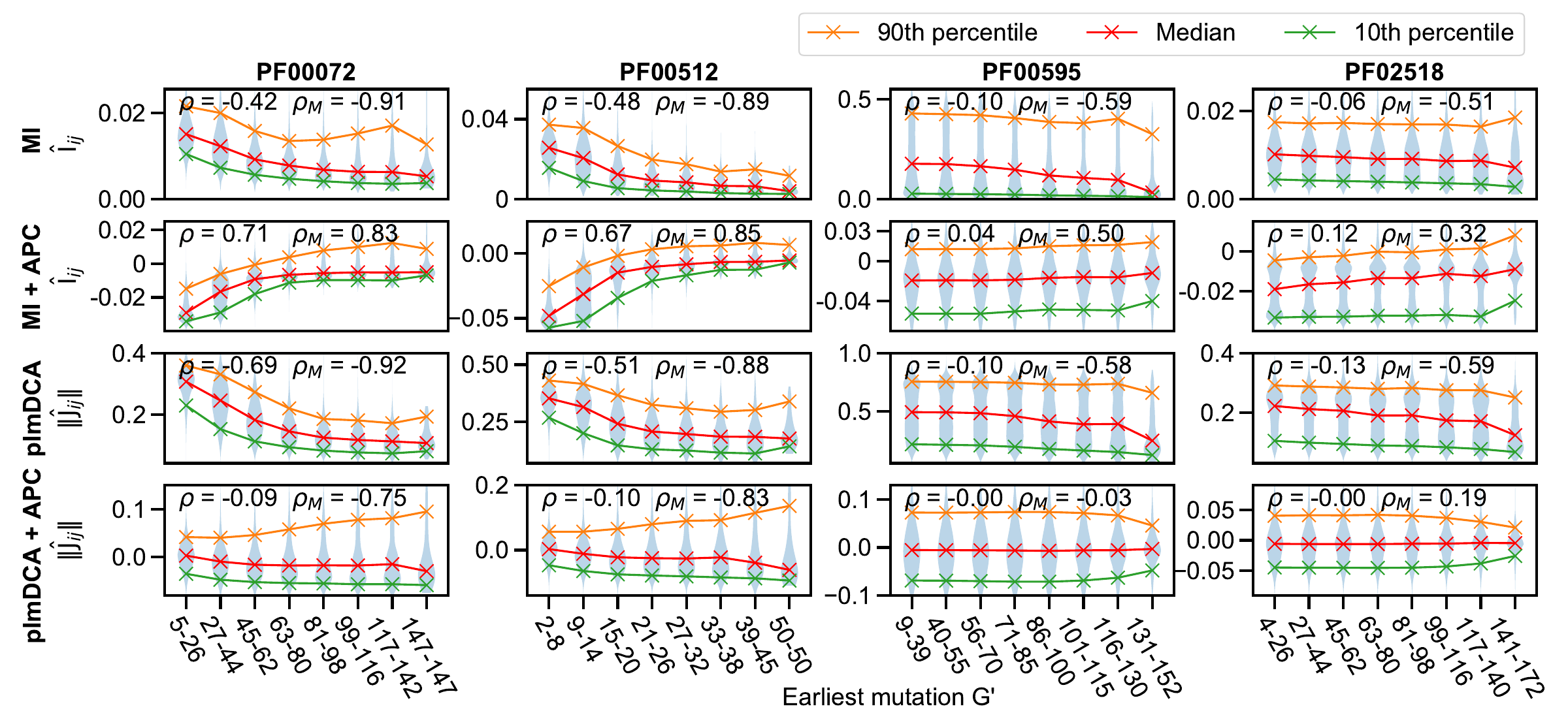}
\caption{\textbf{Impact of phylogeny on coevolution scores for more realistic data.}
Violin plots of the coevolution scores of all pairs of sites $(i,j)$, inferred by MI or plmDCA, without or with APC, are shown versus the earliest mutation $G'$ where both $i$ and $j$ have mutated with respect to the ancestral sequence. To determine $G'$, we count the number $G_i$ (resp.\ $G_j$) of mutations undergone by a sequence when following the branches of the phylogenetic tree from the ancestral sequence to the first one where $i$ (resp.\ $j$) has mutated, and compute $G'=\max(G_i,G_j)$. Pairs $(i,j)$ are binned by the value of $G'$ for the violin plot representation. Two Pearson correlation coefficients are shown: the first one, $\rho$, is obtained when considering all coupling values, and the second one, $\rho_M$, is obtained when considering only the median of the coupling values. Note that no binning on $G'$ is employed in the computation of these Pearson correlations. 
Couplings were inferred on 100 realisations of the data generation in each case, and aggregated together.   }
\label{fig:violinplots_morerealisticdata}
\end{figure}

\paragraph{Extension to more realistic synthetic data, generated from models inferred on natural data.} How well does the insight gained from our minimal model apply to natural data? To address this question, we generate synthetic data directly using DCA Potts models inferred on natural MSAs, and phylogenies inferred on the same MSAs. Specifically, we construct three different data sets for each of the four protein families in Table~\ref{tab:MSA}: natural sequences, equilibrium sequences from the inferred DCA model, and sequences generated using the same inferred DCA model, but along an inferred phylogenetic tree (see Methods). These synthetic data sets are closer to natural data than those generated from our minimal model, but retain the advantage that we know when in the phylogeny each mutation occurred. Figure \ref{fig:violinplots_morerealisticdata} shows the effect of phylogenetic correlations on couplings inferred by mutual information and plmDCA with or without APC on this data. The main conclusion from our minimal model still holds: higher coevolution scores are obtained for pairs of sites that mutate early in the phylogeny, for both MI and plmDCA without APC. The magnitude of this trend depends on the protein family considered, which may be due to the diversity of their phylogenies (note that the ratio of the effective number of sequences to the actual number of sequences takes diverse values among these families, see Table~\ref{tab:MSA}). How do phylogenetic corrections impact this trend? We first compare the effects of different phylogenetic corrections (APC, phylogenetic reweighting) on the performance of contact inference on natural sequences, see Table~\ref{tab:ppv_natseq}. While both corrections enhance performance and their combination is best, APC improves contact prediction substantially more than phylogenetic reweighting. Consistently, Figure \ref{fig:violinplots_morerealisticdata} shows that APC strongly impacts the dependence of coupling strength on phylogeny, as we then generally obtain smaller absolute Pearson correlations of the coevolution scores with the earliest mutation, especially for plmDCA. The overall trend is even reversed by APC for MI, as APC-corrected MI scores are lower for pairs of sites that mutate early in the phylogeny. 

The contact maps for each of these natural and realistic synthetic data sets are shown in Figures \ref{fig:contactmaps_realisticdata_miapc} and \ref{fig:contactmaps_realisticdata_plm}, using MI and plmDCA, respectively, and the TP fractions with or without phylogenetic corrections (phylogenetic reweighting and/or APC) are shown in Table~\ref{tab:ppv_fp_nat_realistic_data}. Note that here, contrarily to our minimal model, the couplings of the inferred DCA models contain some phylogenetic contributions. However, it is fair to compare the equilibrium (independent) and phylogenetic data sets, using the contact map from the bmDCA model (see Methods and~\cite{Figliuzzi18}) as ground truth. We find that phylogeny deteriorates contact prediction performance in these data sets, corroborating the findings from our minimal model.  
Besides, Table~\ref{tab:ppv_fp_nat_realistic_data} confirms that phylogenetic corrections almost always improve the performance of contact prediction. We also find that the number of false positive predicted contacts having shortest path length $L=2$, i.e.\ indirect correlations, is smaller for plmDCA than for MI, consistently with Figure~\ref{fig:HistogramsSPL} (see Table~\ref{tab:ppv_fp_nat_realistic_data}). Finally, and perhaps most interestingly, the number of these FPs with $L=2$ is often smaller in the data set generated with phylogeny than in the equilibrium one, even though the TP fractions decrease as well, suggesting that most false positives are due to phylogeny (and not to the network of contacts) in the phylogenetic data sets.

\section*{Discussion}

Global statistical methods, known as Potts models or DCA, outperform local methods based on covariance or mutual information at the task of unsupervised prediction of structural contacts from protein sequence data~\cite{Weigt09,Marks11,Morcos11}. Perhaps surprisingly, these global methods do not outperform local methods as clearly when applied to synthetic data generated independently at equilibrium from DCA models~\cite{Ngampruetikorn22}. A usual justification for the success of global methods for protein contact prediction is that they allow to disentangle direct and indirect correlations in the data~\cite{Weigt09}, an effect that is present in synthetic data as well, and can be tuned by varying sampling temperature around a ferromagnetic-paramagnetic phase transition~\cite{Gandarilla20,Ngampruetikorn22}. An important difference between natural data and synthetic data sampled independently at equilibrium is the fact that homologous natural sequences share the same ancestry, and thus feature correlations due to phylogeny~\cite{Casari95,Qin18}. These correlations obscure structural ones in the identification of structural contacts both by local~\cite{Dunn08} and global statistical methods~\cite{Qin18,Vorberg18,RodriguezHorta19,RodriguezHorta21}. 

In this context, we investigated the impact of phylogeny on inference performance by generating synthetic data from a minimal model incorporating both structural contacts and phylogeny. We showed that substantial correlations appear between sites that mutate early in the phylogeny, yielding false positive contacts. We found that Potts models are substantially more robust to phylogeny than local methods. This result holds whether or not phylogenetic corrections are used. This robustness to phylogeny contributes to explaining the success of global methods. We showed that our findings from the minimal model generalise well to more realistic synthetic data generated from models inferred on natural data.

The resilience of Potts models to phylogenetic correlations can be understood qualitatively in the simple mean-field approximation, where the DCA couplings are the elements of the inverse covariance matrix of the MSA. Indeed, phylogenetic correlations are associated to large eigenvalues of the covariance matrix~\cite{Casari95,Qin18}, whose overall impact is reduced by inverting the covariance matrix~\cite{Qin18}. 

Our findings illustrate the crucial impact of data structure on inference performance. Protein MSAs are highly structured data sets, with correlations coming from functional constraints such as structural contacts, as well as from phylogeny, and disentangling these signals is fundamentally difficult~\cite{WeinsteinPreprint}. The fact that Potts models succeed better than local methods at this task sheds light on the reasons of their performance for contact inference from natural MSAs. Very recently, protein language models based on MSAs, especially MSA Transformer, have outperformed Potts models at unsupervised contact prediction~\cite{rao2021msa}. We showed in another work that they are even more resilient to phylogenetic noise than Potts model~\cite{LupoPreprint}. Taken together, the present results and those of~\cite{LupoPreprint} demonstrate the crucial importance of disentangling phylogenetic correlations from functional ones for contact prediction performance. The ability to partly disentangle these correlations is one of the reasons of the success of Potts models, as we showed here, and also of protein language models, as we showed in~\cite{LupoPreprint}. While phylogenetic correlations are an issue for structure prediction, they are nevertheless an important and helpful signal for the inference of interaction partners among the paralogs of two protein families~\cite{Marmier19,Gerardos22,GandarillaPreprint}. They are thus a double-edged sword for inference from MSAs. It is our hope that a better understanding of the impact of data structure on inference performance will help developing efficient and interpretable methods in the future.

\section*{Models and methods}

\subsection*{Synthetic data generation}

\paragraph{General approach.} We generate synthetic sequences with fixed structural constraints, modelling those that would exist in a given protein family that needs to maintain structure and function. We consider two types of data sets. The first one is generated at equilibrium under these structural constraints. The second one has an evolutionary history given by a simple phylogeny, and we assume that the same structural constraints have existed throughout this evolutionary history.

To model structural constraints, we use an Erd\H{o}s-Rényi random graph with 200 nodes and edge probability $q = 0.02$. This graph is held fixed because it models the structural constraints of a given protein family. Each node corresponds then to an amino-acid site, and each edge to a structural contact. This number of sites and this density of contacts are in the range of those observed in natural proteins, see Table~\ref{tab:MSA}. However, because contact density depends on the threshold used to define contacts, it is interesting to consider larger values of $q$, see Table~\ref{tab:MSA}. In the Supplementary Material, section \ref{comp}, we also consider a graph with  $q = 0.2$~\cite{Ngampruetikorn22}. For simplicity, we model amino acids as Ising spins, i.e.\ binary variables. We consider that couplings between sites exist on the graph edges, and within our minimal model, we take a unique ferromagnetic coupling value for them (set to one). The corresponding Hamiltonian reads
\begin{equation}
H(\vec \sigma)=-\sum_{i=1}^{\ell}h_{i}\sigma_i - \sum_{j=1}^{\ell}\sum_{i=1}^{j-1}J_{ij}\sigma_i\sigma_j= - \sum_{(i,j)\in {\cal E}} \sigma_i \sigma_j\; ,
\label{miniHam}
\end{equation}
where a sequence $\vec \sigma = (\sigma_1,..,\sigma_{\ell}) \in {\{ \pm 1\}^{\ell}}$ gives the state of each Ising spin, while $\ell=200$ is the sequence length, and $\cal E$ is the set of edges in the Erd\H{o}s-Rényi graph. The parameter $h_{i}$ represents the field at site $i$ (here $h_i = 0$ for all $i$) and $J_{ij}$ the coupling between sites $i$ and $j$ ($J_{ij} = 1$ if $(i,j)\in {\cal E}$, and $0$ otherwise).

\paragraph{Generating sequences with structural constraints only.} Independent equilibrium sequences are generated by Metropolis Monte Carlo sampling using the Hamiltonian in Eq.~\ref{miniHam}, which models structural constraints. Each sequence is randomly and independently initialised. Then, moves are proposed, using one of two different methods, either by flipping single spins or clusters of spins (Wolff cluster algorithm)~\cite{Krauth}. In both methods, these moves are accepted or rejected according to the Metropolis criterion, with probability
\begin{equation}
    p = \min \left[ 1, \exp \left(-\frac{\Delta H}{T} \right)\right],
    \label{MetroCrit}
\end{equation}
where $\Delta H$ is the difference in the value of $H$ after and before the flip, and $T$ is the Monte Carlo sampling temperature.
This process is iterated until a certain number of moves are accepted. We choose this number so that equilibrium is reached, which can be checked by the saturation of the absolute magnetisation, see Figure~\ref{fig:Magnetisation_vs_tau}. Note that the decorrelation time of absolute magnetisation is similar to the magnetisation saturation time.

\paragraph{Generating sequences with structural constraints and phylogeny.}
To include phylogeny in the sequence generation process, we start from an equilibrium ancestor sequence produced as explained above at a given temperature $T$. Then, this sequence is evolved by successive duplication and mutation events (``generations'') on a binary branching tree, for a total of $\mathcal{G} = 11$ generations, see Figure~\ref{fig:figure_conceptual}. Mutations are modelled as single spin flips, and each of them is accepted according to the Metropolis criterion at temperature $T$ (see Eq.~\ref{MetroCrit}) with the Hamiltonian in Eq.\ref{miniHam} accounting for structural constraints. We perform a fixed number $\mu$ of accepted mutations on each branch of the phylogenetic tree. The sequences at the leaves of the phylogenetic tree provide a data set of $2^{\mathcal{G}} = 2^{11} = 2048$ sequences.

Because we start from an ancestral equilibrium sequence, and then employ the Metropolis criterion at the same temperature $T$, all sequences in the phylogeny are equilibrium sequences. Thus, these sequences comprise correlations from the structural constraints, as without phylogeny. Their relatedness adds extra correlations, that we call phylogenetic correlations.

\paragraph{Generating more realistic sequences.} To extend our study to more realistic data, we generated sequences from Potts models inferred on natural sequence data. Specifically, natural sequences (Pfam Full multiple sequence alignments~\cite{mistry2021pfam}) were retrieved for four different protein families (PF00072, PF00512, PF00595, PF02518). Next, a bmDCA Potts model~\cite{Figliuzzi18} was inferred on each of these natural multiple sequence alignments. Indeed, bmDCA has a good generative power, demonstrated experimentally~\cite{Russ20}, and thus allows to generate new sequences by Metropolis MCMC sampling~\cite{Figliuzzi18}. (Adaptive Cluster Expansion, ACE~\cite{Barton16,Bravi20}, an alternative generative method, proved too computationally demanding for our analysis.) We used this procedure to produce an equilibrium data set of sequences from each of the inferred Potts models.
We also generated another set of sequences that contains phylogeny using each of these Potts models~\cite{Gerardos22,LupoPreprint}. For this, we first inferred a phylogenetic tree using FastTree~\cite{Price10} from the natural sequences. At its (arbitrary) root, we place a sequence generated at equilibrium by bmDCA. Then, the sequence evolves according to the inferred phylogeny, where proposed mutations are accepted using the same Metropolis criterion as for the equilibrium data. All sequences at the leaves are collected to form the phylogenetic data set. Thus, three data sets are available for each protein family: a natural data set, an equilibrium synthetic data set, and a phylogenetic synthetic data set.

\subsection*{Inferring contacts from sequences}

\paragraph{General approach.} We employ methods developed for contact prediction in natural proteins to infer the edges (or contacts) in the graph, and investigate how they are affected by phylogeny.
In this spirit, the sequences (either a set of equilibrium sequences generated at a given $T$, or all sequences at the leaves of a given phylogenetic tree) are put together to form a multiple sequence alignment (MSA), i.e.\ a matrix where each row is a sequence and each column is a site. Note that here, there is no alignment issue because each node of the graph is well identified. We compare four different inference methods, which attribute a score to each pair of sites (nodes) in the Erd\H{o}s-Rényi graph. Top scoring pairs are predicted contacts. The first two inference methods are local methods, based respectively on covariance ($C$) and on Mutual Information (MI)~\cite{Dunn08}. The other two methods are global methods, which consist in learning a maximum entropy model consistent with the one- and two-site frequencies observed in the data. These pairwise maximum entropy models, also known as Potts models, or Direct Coupling Analysis (DCA)~\cite{Weigt09}, can be approximately inferred in various ways. We consider two widely used versions, namely mean-field DCA (mfDCA)~\cite{Marks11,Morcos11}, which is the simplest one, and pseudolikelihood maximisation DCA (plmDCA)~\cite{Ekeberg13,Ekeberg14}, which is the state-of-the-art contact prediction method based on Potts models. 

All these methods start from the single-site frequencies of each state $\sigma_i$ at each site (or column in the MSA) $i \in {\{ 1,..,\ell \}}$, denoted by $f_i(\sigma_i)$, and the two-site frequencies $f_{ij}(\sigma_i, \sigma_j)$. In addition, MI and DCA methods employ regularisation schemes. First, a pseudocount is added in the computation of the frequencies for MI~\cite{Bitbol18,Marmier19} and for mfDCA~\cite{Morcos11,Marks11}, preventing divergences when computing MI or when inverting the covariance matrix in mfDCA. The pseudocount-corrected frequencies read 
\begin{align}
\tilde{f}_i(\sigma_i) &= \lambda/2 + (1-\lambda)f_i(\sigma_i)\,,\label{f1}\\ \tilde{f}_{ij}(\sigma_i,\sigma_j) &= \lambda/4 + (1-\lambda)f_{ij}(\sigma_i,\sigma_j) \,\,\,\,\,\mathrm{for}\,\,\,\,\, i \neq j\,,\\
\tilde{f}_{ii} (\sigma_i,\sigma_j) &= \delta_{\sigma_i \sigma_j}\tilde{f}_i(\sigma_i)\,,\label{f3}
\end{align}
where $\delta_{\sigma_i \sigma_j} = 1 $ for $\sigma_i = \sigma_j$ and $0$ otherwise. This correction improves contact prediction from protein sequences by mfDCA~\cite{Morcos11,Marks11}. In plmDCA, no pseudocount is employed, but an $L^2$ norm regularisation is used, with strengths $\lambda_h$ on fields and $\lambda_J$ on couplings~\cite{Ekeberg13,Ekeberg14}. Unless otherwise specified, we use the standard values for all these regularisation parameters, namely $\lambda = 0.01$ for MI, $\lambda = 0.5$ for mfDCA and $\lambda_J = \lambda_h = 0.01$ for plmDCA.  An analysis of the impact of pseudocount and regularisation strength on contact prediction by mfDCA and plmDCA is provided in Figure~\ref{fig:tpfrac_vs_pseudocounts}. We compare results with and without APC~\cite{Dunn08}, but we do not employ phylogenetic reweighting~\cite{Weigt09} in the analysis of our minimal model, because the phylogeny is balanced in that case, which would yield threshold effects in this correction. Besides, we find in our analysis of natural and more realistic data that APC yields a more substantial performance increase than phylogenetic reweighting.

\paragraph{Covariance.} The covariance between two sites $i$ and $j$ reads $C_{ij} = \langle \sigma_i \sigma_j \rangle - \langle \sigma_i \rangle \langle \sigma_j \rangle$, where $\langle \cdot \rangle$ denotes a mean across all sequences of the MSA. While this standard definition can be employed directly with Ising spins, the state of a site in a protein sequence is a categorical random variable with 21 possible states (the 20 natural amino acids and the gap), and therefore frequencies are generally employed instead of means. The two descriptions are equivalent for Ising spins. Indeed, 
\begin{equation}
\langle \sigma_i \rangle = f_{i}(1)- f_{i}(-1) = 2f_{i}(1) -1\,, \label{meani}
\end{equation}
and
\begin{equation}
\langle \sigma_i \sigma_j \rangle = f_{ij}(1,1)  - f_{ij}(1,-1)- f_{ij}(-1,1) + f_{ij}(-1,-1)= 1-2f_i(1)-2f_j(1)+4f_{ij}(1,1)\,, \label{meanij}
\end{equation}
where we have employed the marginalisation relationships, $f_i(\sigma)=f_{ij}(\sigma,1)+f_{ij}(\sigma,-1)$ and $f_j(\sigma)=f_{ij}(1,\sigma)+f_{ij}(-1,\sigma)$ for $\sigma\in\{-1,1\}$, and the normalisation relationship $f_i(1)+f_i(-1)=1$, which yield $f_{ij}(1,-1)=f_i(1)-f_{ij}(1,1)$ and $f_{ij}(-1,1)=f_j(1)-f_{ij}(1,1)$, as well as $f_{ij}(-1,-1)=f_i(-1)-f_{ij}(-1,1)=1-f_i(1)-f_j(1)+f_{ij}(1,1)$. Combining Eqs.~\ref{meani} and~\ref{meanij} yields 
\begin{equation}
C_{ij}=\langle \sigma_i \sigma_j \rangle -\langle \sigma_i \rangle \langle \sigma_j \rangle  = 4\left[f_{ij}(1,1) - f_i(1)f_j(1)\right]\,.
\end{equation}
We employ the absolute value $|C_{ij}|$ of the covariance to score each pair of sites $(i,j)$. 

\paragraph{Mutual information.} The MI of each pair of sites $(i,j)$ is computed as
\begin{equation}
\mathrm{MI}_{ij} = \sum_{\sigma_i, \sigma_j} \tilde{f}_{ij}(\sigma_i, \sigma_j) \log\left( \frac{\tilde{f}_{ij}(\sigma_i, \sigma_j)}{\tilde{f}_i(\sigma_i) \tilde{f}_j(\sigma_j) }\right),
\label{mi_eq}
\end{equation}
where $\tilde{f}_i(\sigma_i)$ and $\tilde{f}_{ij}(\sigma_i,\sigma_j)$ are the pseudocount-corrected one- and two-body frequencies (see Eqs.~\ref{f1}-\ref{f3}). Note that we use frequencies instead of probabilities to estimate mutual information, and do not correct for finite-size effects~\cite{Bialek}. Indeed, we only compare scores computed on data sets with a given size, affected by the same finite size effects.

\paragraph{Potts models (DCA).} DCA methods aim to infer the pairwise maximum entropy probability distribution of the data, which is the least constrained distribution consistent with the empirically measured one- and two-body frequencies. The probability of a sequence $\vec{\sigma}$ has the following form:
\begin{equation}
\label{maxentropy_eq}
    P(\vec{\sigma}) = \frac{\exp\left[ -H (\vec{\sigma}) \right]}{Z},
\end{equation}
where the Hamiltonian $H$ is given by Eq.~\ref{miniHam}, i.e.\ the Potts model Hamiltonian, and $Z$ is a normalisation constant (partition function). The mfDCA method provides a simple approximation of the couplings $J_{ij}$ in the zero-sum (Ising) gauge~\cite{Morcos11,Marks11} given by $\hat{J}_{ij} = -\left( \tilde{C}^{-1} \right)_{ij}$, where $\tilde{C}$ is the covariance matrix computed with the pseudocount-corrected frequencies (see Eqs.~\ref{f1}-\ref{f3}), whose elements read $\tilde{C}_{ij} = (1-\lambda)\left[ C_{ij} + \lambda \langle \sigma_i \rangle \langle \sigma_j \rangle \right] = (1-\lambda)\langle \sigma_i \sigma_j \rangle-(1-\lambda)^2 \langle \sigma_i \rangle \langle \sigma_j \rangle$ for $i \neq j$ and $\tilde{C}_{ii} = (1-\lambda)^2 C_{ii} + \lambda(2-\lambda)=(1-\lambda)^2(1-\langle \sigma_i\rangle^2) + \lambda(2-\lambda)$. Finally, the plmDCA method provides another estimation $\hat{J}_{ij}$ of the couplings $J_{ij}$ by maximising the pseudolikelihood of Eq.~\ref{maxentropy_eq}. We also use the zero-sum gauge for plmDCA~\cite{Ekeberg13,Ekeberg14}, and we employ the absolute value  $|\hat{J}_{ij}|$ of the inferred coupling to score each pair of sites $(i,j)$.  With natural sequences and realistic synthetic sequences with 21 states, the Frobenius norm of the $21\times 21$ matrix of coupling values $\hat{J}_{ij}(\alpha,\beta)$ is used to obtain one score per pair of site~\cite{Morcos11,Marks11} and it is denoted by $\Vert \hat{J}_{ij} \Vert$. 

\paragraph{Evaluating performance.} We assess performance via the fraction of correctly identified contacts among the $N_{\mathrm{contacts}}$ top-scoring pairs of sites, where $N_{\mathrm{contacts}}=413$ is the number of actual contacts, i.e.\ of edges in the Erd\H{o}s-Rényi random graph used to generate the data. We refer to this fraction as the True Positive fraction (TP fraction). Because we evaluate it for a number of predicted contacts equal to the number of actual contacts, we have $\textrm{TP}+\textrm{FP}= \textrm{TP}+\textrm{FN} = N_{\mathrm{contacts}}$, where TP denotes the true positives, FP the false positives and FN the false negatives. Thus, the TP fraction is equal to the sensitivity, recall, hit rate or true positive rate $ \mathrm{TPR} = \textrm{TP}/(\textrm{TP}+\textrm{FN})$, but also to the precision or positive predictive value $ \mathrm{PPV} =  \textrm{TP}/(\textrm{TP}+\textrm{FP})$.

We employ the TP fraction (or PPV) as a measure of performance because the usual practice for natural sequence data is to predict the top scoring pairs of sites as contacts. 
It is also interesting to evaluate how the quality of predictions depends on the number of predicted contacts. For this, one can consider the area under the receiver operating characteristic (AUC). We found very similar conclusions for the AUC as for the TP fraction, see Figures~\ref{fig:TPAUC_vs_T_densegraph_eqm15} and \ref{fig:TPAUC_vs_T_densegraph_m5m15}.

\paragraph{Extension to more realistic data.} Inference on natural and more realistic data is performed using one local method and one global method, respectively MI and plmDCA.
For MI, we employ Eq.~\ref{mi_eq}, using a pseudocount of 0.01, with $\sigma_i \in \{1,...,21\}$. For plmDCA, we use the implementation of the \texttt{PlmDCA} package (\href{url}{https://github.com/pagnani/PlmDCA}). Unless otherwise specified, the default values of the parameters of this package are used.

For natural and more realistic data, the number of predicted contacts is taken as $N_{\textrm{pred}} = 2\ell$, where $\ell$ is the length of the protein (number of amino-acid sites, reported in Table~\ref{tab:MSA}), and performance is measured as the PPV (or TP fraction) at this number of predicted contacts~\cite{Morcos11,Marks11}. 

\section*{Code accessibility}
Our code is freely available at \href{url}{https://zenodo.org/record/7503931}.

\section*{Acknowledgements}

N.\ D.\ and A.-F.\ B. thank Matthieu Wyart for helpful discussions. This project has received funding from the European Research Council (ERC) under the European Union’s Horizon 2020 research and innovation programme (grant agreement No. 851173, to A.-F.\ B.).

\clearpage
\newpage

\begin{center}
 {\LARGE \bf Supplementary material}   
\end{center}

\renewcommand{\thesection}{S\arabic{section}}
\renewcommand{\thefigure}{S\arabic{figure}}
\setcounter{figure}{0}
\renewcommand{\thetable}{S\arabic{table}}
\setcounter{table}{0} 

\tableofcontents

\section{Data generation at equilibrium in the minimal model}

\begin{figure}[htbp]
\centering
\includegraphics[scale = 0.4]{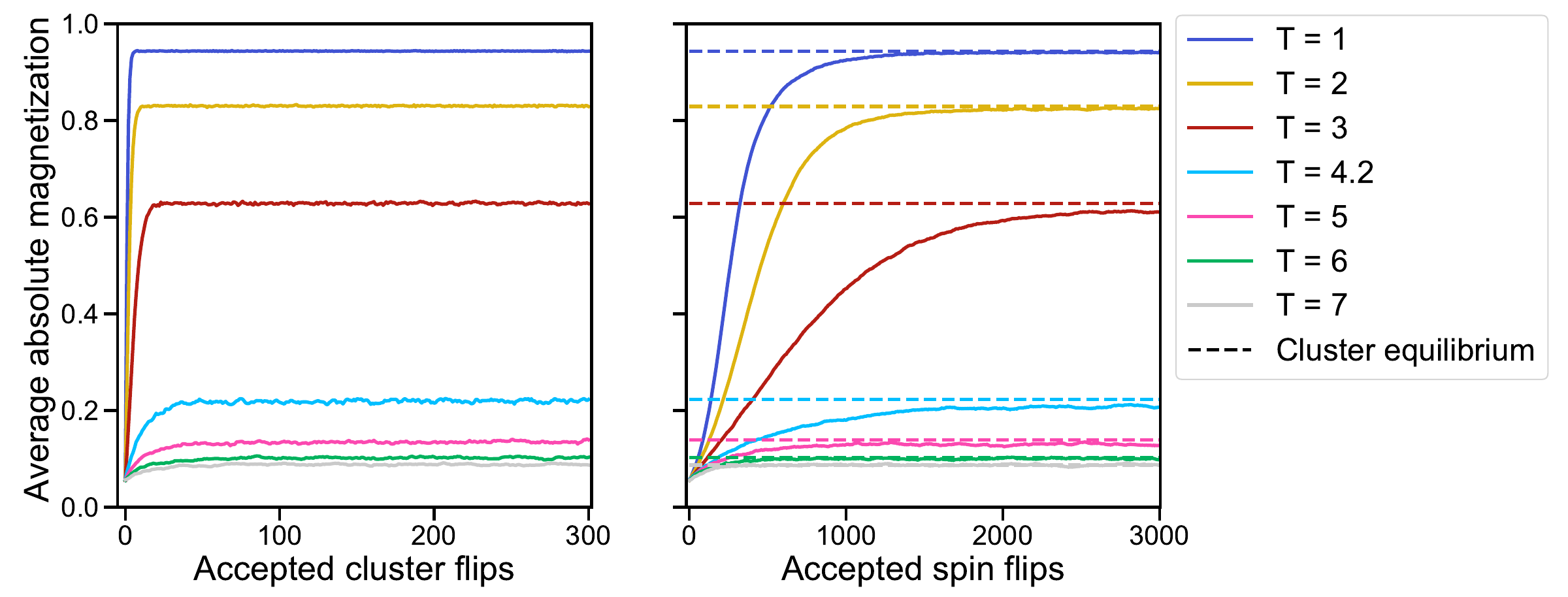}
\caption{\textbf{Generation of independent equilibrium sequences.} The average absolute magnetisation per site is shown versus the number of accepted cluster flips (left panel) or spin flips (right panel) during Metropolis Monte Carlo sampling. Initialisation is made from sequences where each spin is chosen uniformly at random. Then, either a cluster (Wolff) algorithm  (left panel) or a single flip algorithm (right panel) is used. Gradually, equilibrium is reached, and magnetisation stabilises on a plateau. This happens faster with the cluster algorithm. The final value reached with this algorithm is shown on the right panel for comparison. The same contact map (Erd\H{o}s-Rényi graph) as in Figure \ref{fig:tpfrac_vs_mu_sparse} is used, and different
sampling temperatures $T$ are employed. The absolute magnetisation is averaged over all sites in the sequence, and over 2048 sequences, each starting from a different random initialisation.}
\label{fig:Magnetisation_vs_tau}
\end{figure}

%\clearpage
\newpage
\section{Impact of small nonzero couplings}
\begin{figure}[htbp]
\centering
\includegraphics[scale = 0.4]{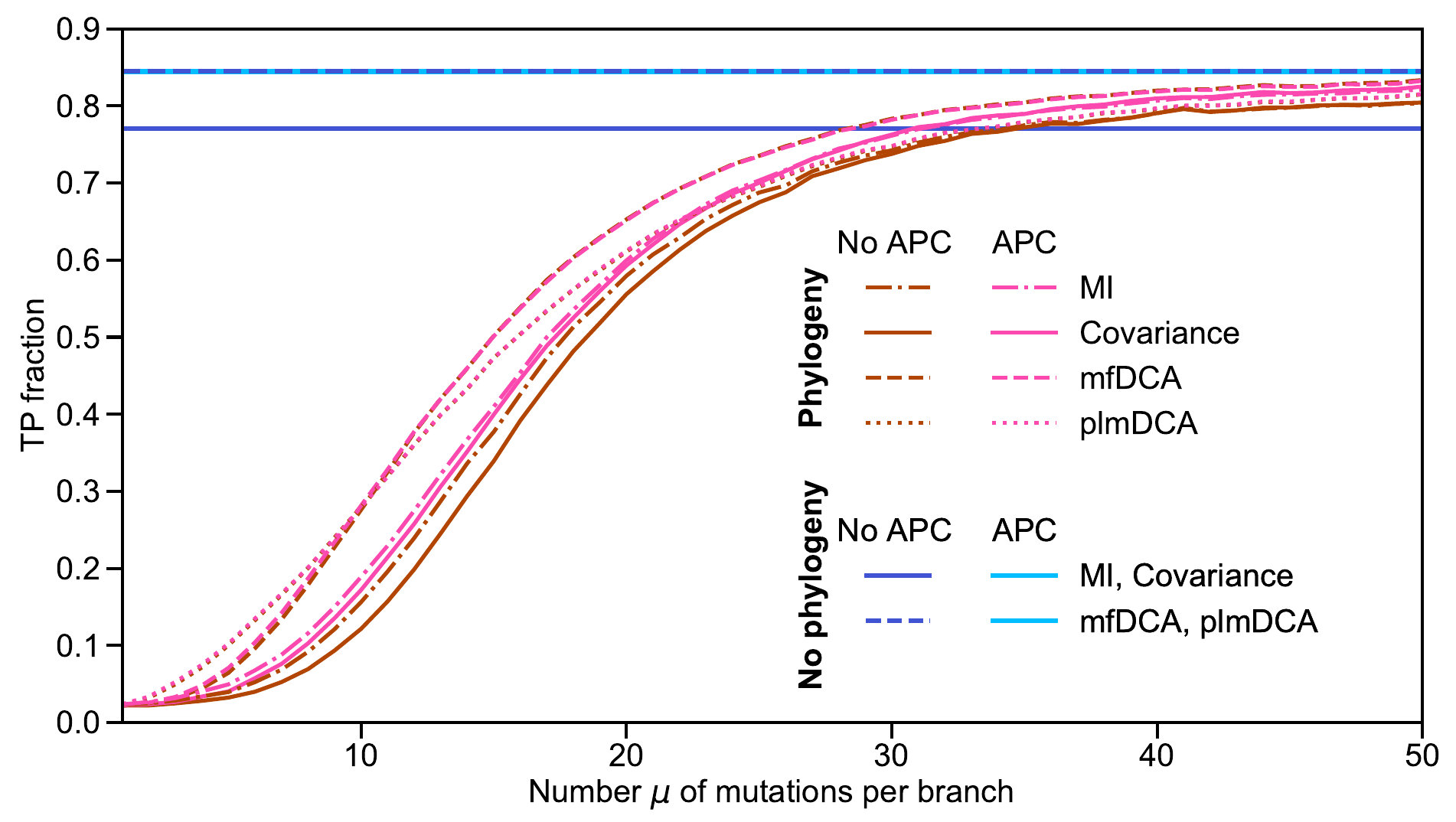}
\caption{ \textbf{Impact of phylogeny on contact prediction with small nonzero couplings.} The TP fraction is plotted versus the number $\mu$ of mutations per branch of the phylogenetic tree, as in Figure~\ref{fig:tpfrac_vs_mu_sparse}. The difference is that here, couplings are distributed according to a Gaussian distribution with mean 0 and standard deviation 1, and then multiplied by $-1$ if negative, to have only positive couplings as in our minimal model. The number of sites is 200 as usual, but a new Erd\H{o}s-Rényi graph is generated with $1300$ contacts, so that there are 414 couplings greater than or equal to 1, which are considered as the actual contacts to be inferred. This number is very close to the number of contacts $N_{\mathrm{contacts}}=413$ in the sparse graph of Figure~\ref{fig:tpfrac_vs_mu_sparse}. However, there are many small nonzero couplings in addition. All results are averaged over $100$ realisations. Note that the performance of local methods with phylogeny and no APC does not tend to that of the corresponding equilibrium case at high $\mu$. This is due to the sampling effect described in section~\ref{conservlowT}.}
\label{fig:tpfrac_vs_mu_gaussianJij}
\end{figure}
\newpage

\section{Impact of conservation on local methods at low $T$}
\label{conservlowT}

\begin{figure}[htbp]
\centering
\includegraphics[scale = 0.35]{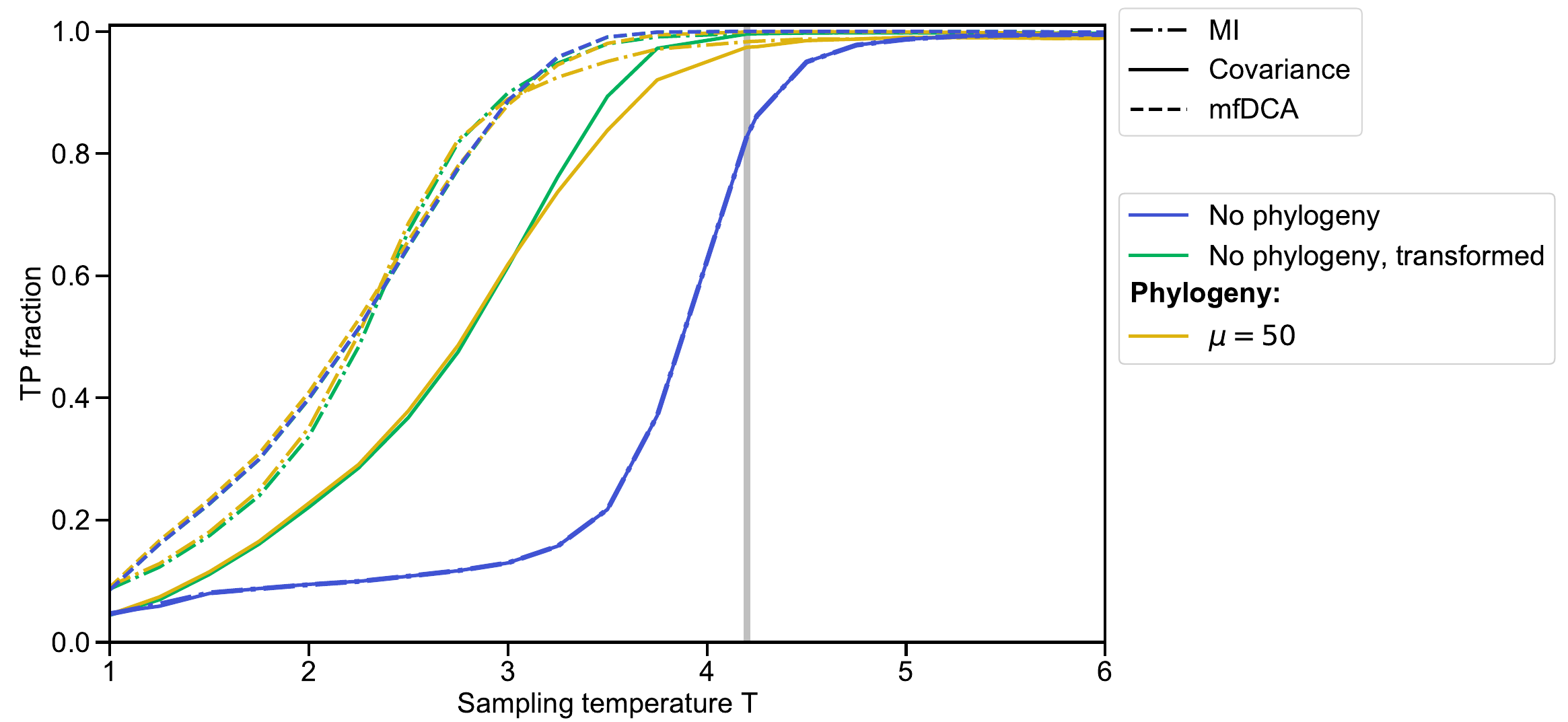}
\caption{\textbf{Impact of data sampling details on contact prediction.} The correctly predicted fraction of contacts (TP fraction) is plotted versus the Monte Carlo sampling temperature $T$ for three different inference methods (Covariance, MI, mfDCA). Note that, for the sake of readability, plmDCA results are not shown, but they behave very similarly as with mfDCA. The equilibrium (no phylogeny) data set employed is the same as in Figure \ref{fig:tpfrac_vs_mu_sparse}. We also consider a transformed data set, built by multiplying by $-1$ the sequences of this equilibrium data set that have negative magnetisation. The data set for $\mu = 50$ is generated as in Figure \ref{fig:tpfrac_vs_mu_sparse}, using the same contact map (Erd\H{o}s-Rényi graph), but with $\mu=50$ (weak phylogeny).}
\label{fig:TP_vs_T_transfdata}
\end{figure}

A surprising feature in the left panel of Figure \ref{fig:tpfrac_vs_T_sparse} is that the performance of local methods is worse at equilibrium than for $\mu = 5$ and $\mu = 15$ at low $T$. Besides, this effect is corrected by the APC (Figure \ref{fig:tpfrac_vs_T_sparse}, right panel). To understand this, let us examine the way the data is generated. At low $T$, deep in the ferromagnetic phase, equilibrium sequences mainly comprise aligned spins, with a large majority of either $1$ (positive overall magnetisation) or $-1$ (negative overall magnetisation). On average, half of the independent equilibrium sequences are in each of these classes, leading to very strongly correlated baseline scores for every pair of sites. In contrast, when generating data with phylogeny at low $T$, all sequences generally have the same overall magnetisation sign, because they all stem from the same equilibrium ancestor, and its magnetisation sign is generally preserved through the phylogeny. The resulting high conservation leads to a baseline of weakly correlated sites, which may allow to detect some signal from the few sites which flipped in the phylogeny, allowing to predict a few contacts. To test this hypothesis, we transform the equilibrium data sets at $T<T_C$ by fully flipping each sequence with negative magnetisation, yielding only sequences with positive magnetisation. Figure \ref{fig:TP_vs_T_transfdata} shows that inference employing local methods is indeed drastically improved by this transformation, while DCA results are very little impacted, probably partly thanks to the large pseudocount in mfDCA and the regularisation in plmDCA. Indeed, a pseudocount can reduce the value of very high correlations and can also turn a strong conservation into a correlation (see details below), thereby somewhat easing the difficulties encountered for these extreme cases. This data transformation further allows to make a direct link between equilibrium data and data generated with little phylogeny: Figure \ref{fig:TP_vs_T_transfdata} shows that data generated with $\mu=50$ behaves similarly as transformed equilibrium data. Importantly, with this transformation, the difference of performance between local and DCA methods for equilibrium data is substantially reduced. In Figure \ref{fig:TP_vs_T_transfdata}, inference using mutual information is almost as good as using DCA for low temperatures. The impact of our data transformation modifying magnetisation signs is in very similar to that of the APC shown in Figure~\ref{fig:tpfrac_vs_T_sparse}. This makes sense, as the APC essentially increases contrast in the score matrices by subtracting the product of means over rows and columns, and can thus partly amend the collective effects observed in equilibrium ferromagnetic sequences. Interestingly, these low-$T$ results show that conservation can help contact inference in some cases.

\paragraph{Impact of the pseudocount in simple cases.} Considering a sequence with just two sites, the covariance matrix can be written as:
\begin{equation}
    C = \begin{pmatrix}
C_{11} & C_{12}\\
C_{21} & C_{22}
\end{pmatrix},
\end{equation}
where $C_{ij} = \langle \sigma_i \sigma_j \rangle - \langle \sigma_i \rangle \langle \sigma_j \rangle$.
Adding a pseudocount $\lambda$ amounts to replacing $C_{ij}$ by:
\begin{align}
\label{Ctilde_ij}
    \tilde{C}_{ij} &= (1-\lambda) \left[ C_{ij} + \lambda \langle \sigma_i \rangle \langle \sigma_j \rangle \right] \,\,\textrm{for}\,\, i \neq j\,,\\
\label{Ctilde_ii}
    \tilde{C}_{ii} &= (1-\lambda)^2\left[1- \langle \sigma_i \rangle^2 \right] + \lambda (2-\lambda)\,.
\end{align}
In the mean-field approximation, the couplings $J$ can be computed from $\tilde{C}$ via $J = -\tilde{C}^{-1}$. In our two-site case, this yields:
\begin{equation}
  \tilde{C}^{-1} = \frac{1}{\mathrm{det}\tilde{C}}\begin{pmatrix}
\tilde{C}_{22} & -\tilde{C}_{12}\\
-\tilde{C}_{21} & \tilde{C}_{11}
\end{pmatrix},
\end{equation}
with 
\begin{equation}
    \mathrm{det}\tilde{C} = \tilde{C}_{11}\tilde{C}_{22}-\tilde{C}_{12}\tilde{C}_{21}\,.\label{det}
\end{equation}
Thus, for instance,
\begin{equation}
    J_{12} = \frac{1}{\mathrm{det}\tilde{C}}\tilde{C}_{12}
\end{equation}
Therefore, in this two-site case, the couplings are equal to the pseudocount-corrected covariance up to the inverse determinant prefactor.

Let us now consider two specific extreme situations in this two-site case. In the first one, $\sigma_1=\sigma_2 = 1$ in every sequence (perfect conservation), while in the second one, $\sigma_1=\sigma_2 = 1$ in one half of the sequences and  $\sigma_1=\sigma_2 = -1$ in the other half of the sequences (perfect correlation, but no conservation).

The first situation yields $C^{(1)}_{12} = 0$ and $C^{(1)}_{11}=0$, leading to a non-invertible covariance matrix without a pseudocount. Including a pseudocount gives $\tilde{C}^{(1)}_{12} = \tilde{C}^{(1)}_{21}= \lambda(1-\lambda)$ (see Eq.~\ref{Ctilde_ij}), which has a maximum at $\lambda = 0.5$, the usual value used in mfDCA. To obtain the couplings, we compute the determinant using Eq.~\ref{det} (and Eq.~\ref{Ctilde_ii}, which provides $\tilde{C}^{(1)}_{11} = \tilde{C}^{(1)}_{22} = \lambda(2-\lambda)$).
Thus,
\begin{equation}
    \text{det}\tilde{C}^{(1)} = \left[\lambda(2-\lambda)\right]^2-\left[\lambda(1-\lambda)\right]^2 = \lambda^2 \left(3-2\lambda \right).
\end{equation}
This gives
\begin{equation}
    J_{12}^{(1)} = \frac{1}{\mathrm{det}\tilde{C}^{(1)}}\tilde{C}_{12}^{(1)} = \frac{1-\lambda}{\lambda \left(3-2\lambda\right)}.
\end{equation}
%yielding $J_{12}$ = $1/2$ for $\lambda  = 0.5$.

In the second situation, $C_{12}^{(2)} = 1$ and $C_{11}^{(2)}=1$, yielding a non-invertible covariance matrix in the absence of a pseudocount. Including a pseudocount gives $\tilde{C}_{12}^{(2)} = 1-\lambda$, while $\tilde{C}_{11}^{(2)} = \tilde{C}_{22}^{(2)} = (1-\lambda)^2 + \lambda(2-\lambda)=1$. The determinant reads
\begin{equation}
    \text{det}\tilde{C}^{(2)} = 1-\left(1-\lambda\right)^2= \lambda(2-\lambda)\,.
\end{equation}
This leads to
\begin{equation}
    J_{12}^{(2)} = \frac{1}{\mathrm{det}\tilde{C}^{(2)}}\tilde{C}_{12}^{(2)} = \frac{1-\lambda}{\lambda \left(2-\lambda\right)}.
\end{equation}
%yielding $J_{12}$ = $2/3$ for $\lambda  = 0.5$.

To compare these two situations, let us first consider the ratio of covariances between the two sites
\begin{equation}
    \frac{\tilde{C}_{12}^{(1)}}{\tilde{C}_{12}^{(2)}} = \frac{\lambda(1-\lambda)}{1-\lambda}=\lambda\,:
\end{equation}
we have $0<\tilde{C}_{12}^{(1)} <\tilde{C}_{12}^{(2)}<1$ for all values of $\lambda \in \left]0,1\right[$. Besides, while in the first situation the two sites have zero covariance in the absence of a pseudocount, we see here that the ratio $\tilde{C}_{12}^{(1)}/\tilde{C}_{12}^{(2)}$ is equal to $1/2$ for the usual choice $\lambda=0.5$ (and even becomes 1 for $\lambda\to 1$).
Let us now turn to the couplings between these two sites. Their ratio is
\begin{equation}
    \frac{J_{12}^{(1)}}{J_{12}^{(2)}} = \frac{1-\lambda}{\lambda \left(3-2\lambda\right)} \frac{\lambda \left(2-\lambda\right)}{1-\lambda}=\frac{2-\lambda}{3-2\lambda}\,:
\end{equation}
we have $0<J_{12}^{(1)} <J_{12}^{(2)}$ for all values of $\lambda \in \left]0,1\right[$. More precisely, the ratio $J_{12}^{(1)}/J_{12}^{(2)}$ increases with $\lambda$, being equal to $2/3$ for $\lambda\to 0$ and to $1$ for $\lambda\to 1$, and to $3/4$ for the usual choice $\lambda=0.5$.

Therefore, using a pseudocount transforms the fully-conserved case from no covariance and an indetermination for the coupling to a covariance value and a coupling value that are finite and comparable to the fully-correlated but not-conserved case. The magnitude of these effective covariances and couplings coming from conservation is larger if the value of $\lambda$ is increased. In particular, for the usual value $\lambda=0.5$, the effective coupling represents $3/4$ of those coming from conservation-free correlation.

%\newpage

\section{Comparison with Ref.~\cite{Ngampruetikorn22}}
\label{comp}

\begin{figure}[h!]
\centering
\includegraphics[scale = 0.4]{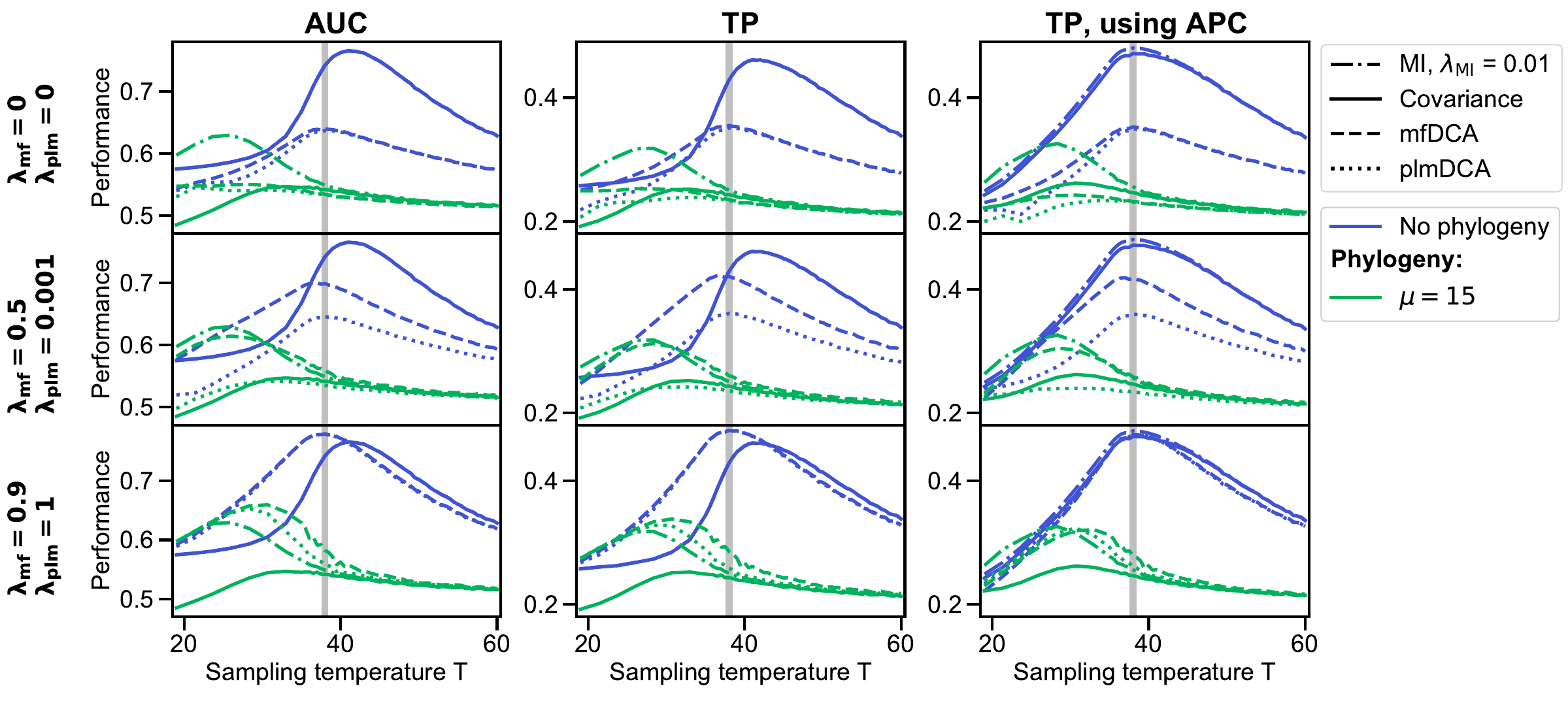}
\caption{\textbf{Contact prediction performance for a denser random graph on equilibrium and phylogenetic data.}
The results of different performance assessments, namely the area under the receiver operating characteristic curve (AUC), the TP fraction and the TP fraction using APC for four inference methods (C, MI, mfDCA, plmDCA) is shown on two different data sets. Data has been generated at equilibrium (no phylogeny) and with phylogeny (here $\mu = 15$) in the same way as in Figure \ref{fig:tpfrac_vs_T_sparse} but using a denser contact map (Erd\H{o}s-Rényi graph with probability $q = 0.2$ instead of $q=0.02$). Each line of plots in this figure represent the inference with different values of pseudocounts and regularisation parameters used in the DCA methods, the actual values are shown on the left in bold. The value of the pseudocount is fixed for MI and none is used for covariance. The top left panel recovers the result found in~\cite{Ngampruetikorn22}. }
\label{fig:TPAUC_vs_T_densegraph_eqm15}
\end{figure}

\begin{figure}[h!]
\centering
\includegraphics[scale = 0.4]{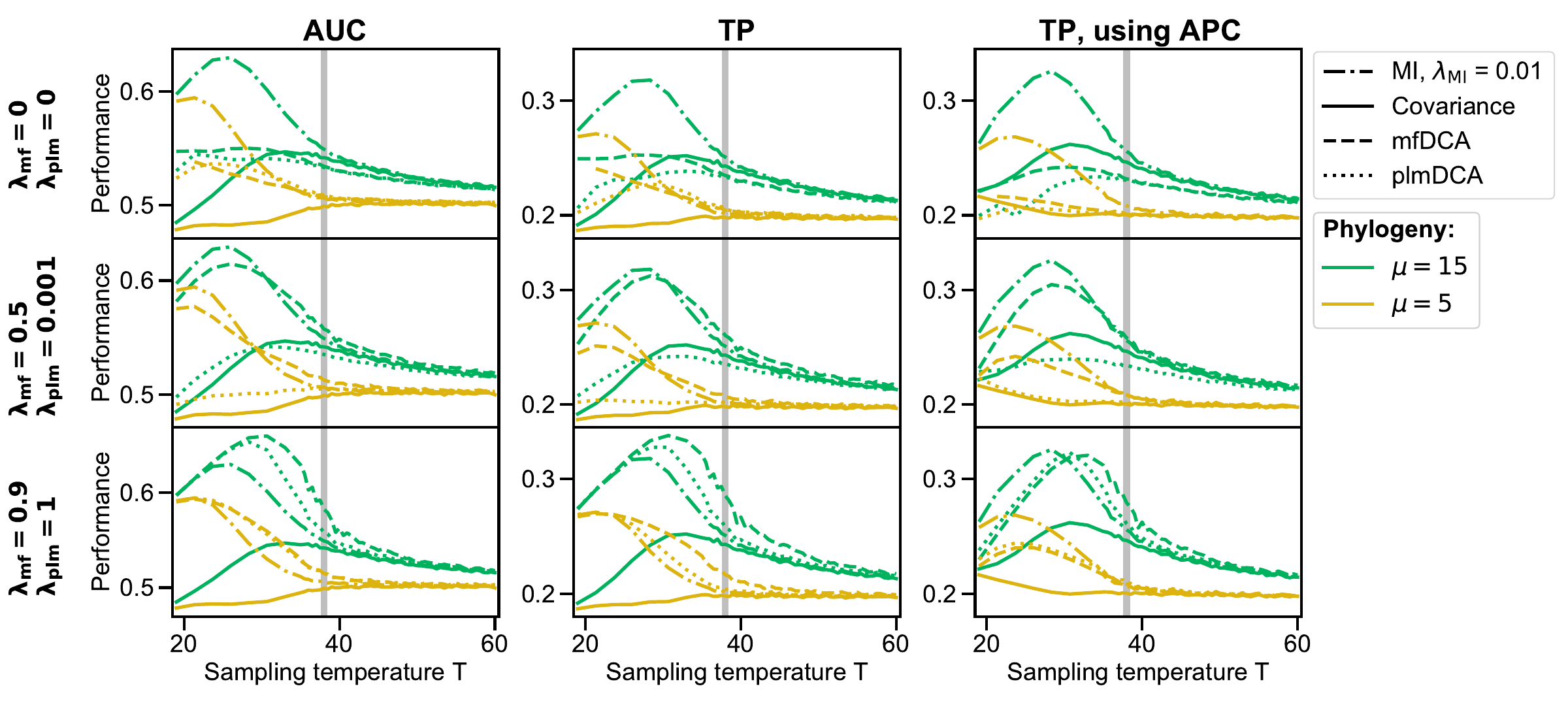}
\caption{\textbf{Contact prediction performance for a denser random graph on two different phylogenetic data sets.}
Same plots as in Figure \ref{fig:TPAUC_vs_T_densegraph_eqm15}, but for two different phylogenetic data sets, one with $\mu = 5$ and the other one with $\mu = 15$.}
\label{fig:TPAUC_vs_T_densegraph_m5m15}
\end{figure}

\begin{figure}[htbp]
\centering
\includegraphics[scale = 0.35]{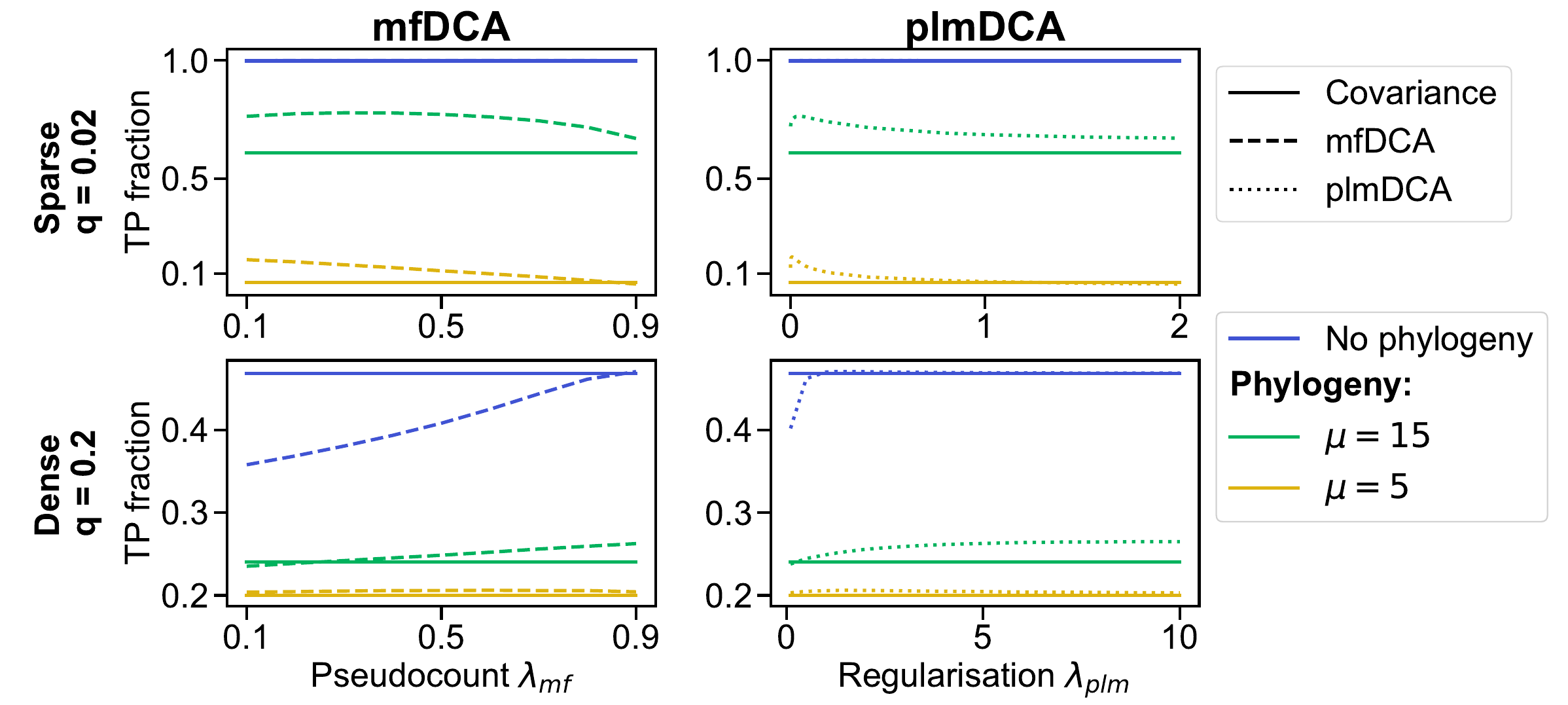}
\caption{ \textbf{Impact of pseudocount and regularisation on contact prediction in sparse and denser graphs.} The fraction of correctly predicted contacts (TP fraction) is plotted versus the pseudocount for mfDCA, or the regularisation strength for plmDCA, in Erd\H{o}s-Rényi graphs with two different probabilities of contact ($q = 0.02$ and $q = 0.2$). In each case, the performance of contact prediction by the covariance matrix is also shown for reference. Three data sets are presented in each case: data generated independently at equilibrium (no phylogeny), and data generated with phylogeny at $\mu = 15$ and $\mu = 5$. For the sparse graph, data is generated as in Figure~\ref{fig:tpfrac_vs_mu_sparse} at $T = 5$. For the denser graph, data is generated as in Figure~\ref{fig:TPAUC_vs_T_densegraph_eqm15} at $T = 40$. All results are averaged over 100 realisations.}
\label{fig:tpfrac_vs_pseudocounts}
\end{figure}

In Ref.~\cite{Ngampruetikorn22}, the performance of mfDCA and covariance was compared on equilibrium data sets generated in a minimal model similar to ours, at different sampling temperatures. However, in Ref.~\cite{Ngampruetikorn22}, covariance was found to perform better than mfDCA for all values of $T$ in data sets of similar size as ours. One difference is that a substantially denser graph was employed. Thus, to make a more in-depth comparison with Ref.~\cite{Ngampruetikorn22}, we generated data using an Erd\H{o}s-Rényi graph with $q = 0.2$ (recall that $q=0.02$ in the rest of our work, $q$ being the probability that two nodes are connected in the graph). Moreover, we started by using no pseudocounts and assessing contact prediction performance via the AUC (area under the receiver operating characteristic), as in Ref.~\cite{Ngampruetikorn22}. Consistently with Ref.~\cite{Ngampruetikorn22}, we find that covariance and mutual information then perform better than DCA for equilibrium data (see top left panels of Figures \ref{fig:TPAUC_vs_T_densegraph_eqm15} and \ref{fig:TPAUC_vs_T_densegraph_m5m15}).
The same result holds when using TP fraction to assess performance (see top middle panels of Figures \ref{fig:TPAUC_vs_T_densegraph_eqm15} and \ref{fig:TPAUC_vs_T_densegraph_m5m15}). Note that TP fraction is most often used in the DCA field, which is why we employ it here, but that the AUC has the advantage of not depending on a threshold (here, a number of predicted contacts).
While these results are obtained without regularisation, as in Ref.~\cite{Ngampruetikorn22}, we observe that high pseudoucount or regularisation strength values allow DCA to reach better prediction performance and outperform covariance and mutual information for low temperatures, while performing similarly for larger temperatures. Interestingly, even higher regularisation strengths are required for this denser graph (see Figures \ref{fig:TPAUC_vs_T_densegraph_eqm15} and \ref{fig:TPAUC_vs_T_densegraph_m5m15}). This illustrates the strong importance of regularisation for the performance of DCA. A more detailed analysis of the impact of pseudocount or regularisation strength on the inference of contacts by DCA is shown in Figure~\ref{fig:tpfrac_vs_pseudocounts}. Hence, our results are consistent with those of Ref.~\cite{Ngampruetikorn22}, and the apparent differences arise mainly from regularisation. Furthermore, Figures \ref{fig:TPAUC_vs_T_densegraph_eqm15} and \ref{fig:TPAUC_vs_T_densegraph_m5m15} confirm that global methods (DCA) handle phylogeny better than local ones (covariance and mutual information) for this denser graph, as for our sparse graph (see Fig.~\ref{fig:tpfrac_vs_T_sparse}). We also studied the impact of APC on inference in the denser graph (see Figures \ref{fig:TPAUC_vs_T_densegraph_eqm15} and \ref{fig:TPAUC_vs_T_densegraph_m5m15}), and found that it slightly deteriorates the inference performance of global methods, especially on data containing phylogeny, while it improves that of local methods, especially on equilibrium data, consistently with our results above.

\newpage

\clearpage
\newpage

\section{Impact of pseudocount and regularisation on contact prediction}

\newpage
\section{Impact of graph properties on coevolution scores}

\begin{figure}[h!]
\centering
\includegraphics[scale = 0.4]{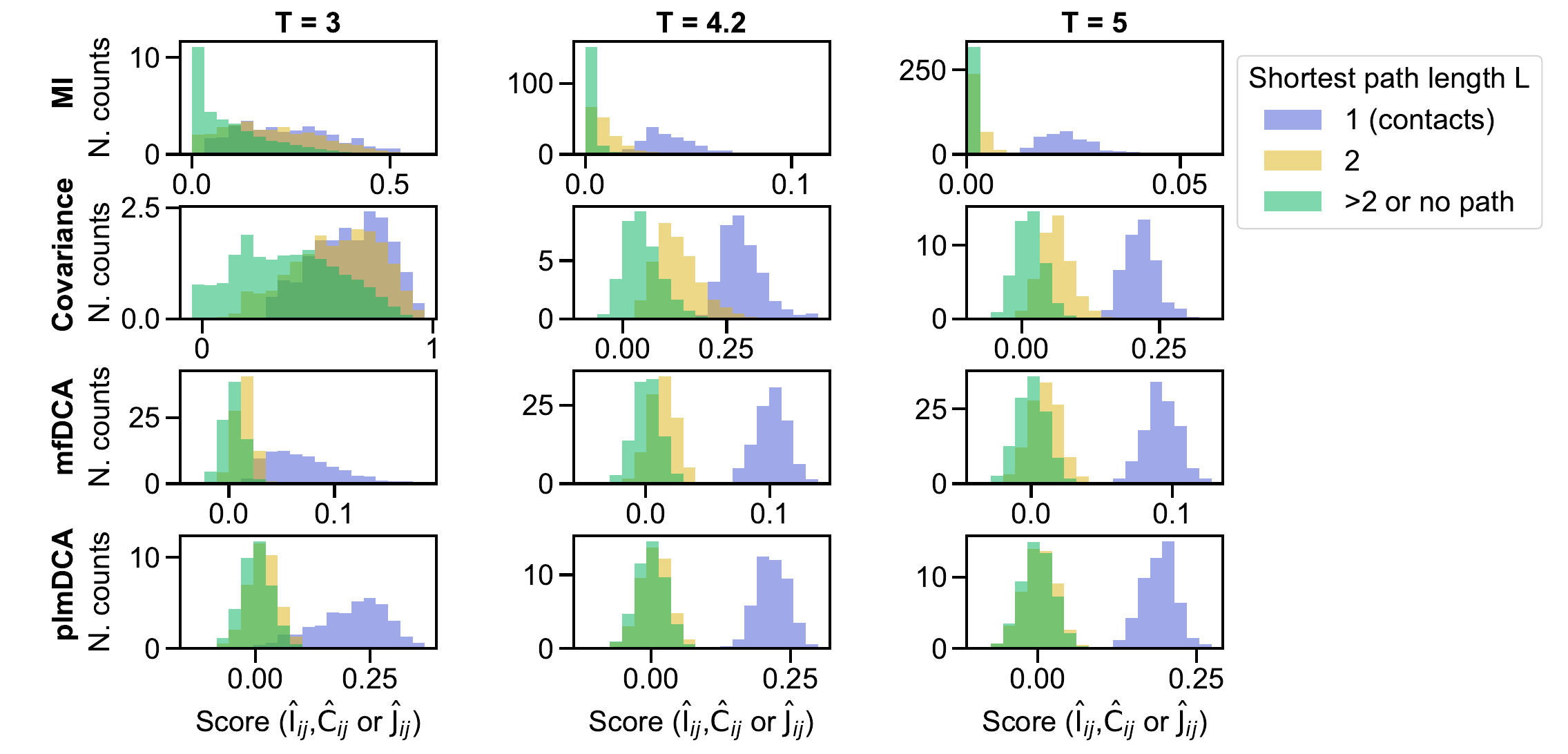}
\caption{\textbf{Impact of graph properties on coevolution scores for equilibrium sequences.} Histograms of coevolution scores for all pairs of sites $(i,j)$ are shown for equilibrium data sampled at three different temperatures: $T=3<T_C$ (left), $T=4.2\simeq T_C$ (center), and $T=5>T_C$ (right), for four inference methods (MI, Covariance, mfDCA, plmDCA) without APC. Pairs of sites are split into three categories according to the length $L$ of the shortest path connecting them in the graph: contacts, $L = 1$; first indirect neighbours, $L = 2$; more distant ($L > 2$ or isolated sites). Normalised counts (N. counts) are shown -- note that there are many more non-contact than contact pairs. Data sets of 2048 sequences each are generated at equilibrium using the cluster algorithm (see Figure \ref{fig:Magnetisation_vs_tau}), using the same contact map (Erd\H{o}s-Rényi graph) as in Figure~\ref{fig:tpfrac_vs_mu_sparse}.}
\label{fig:HistogramsSPL}
\end{figure}

\begin{figure}[h!]
\centering
\includegraphics[scale = 0.4]{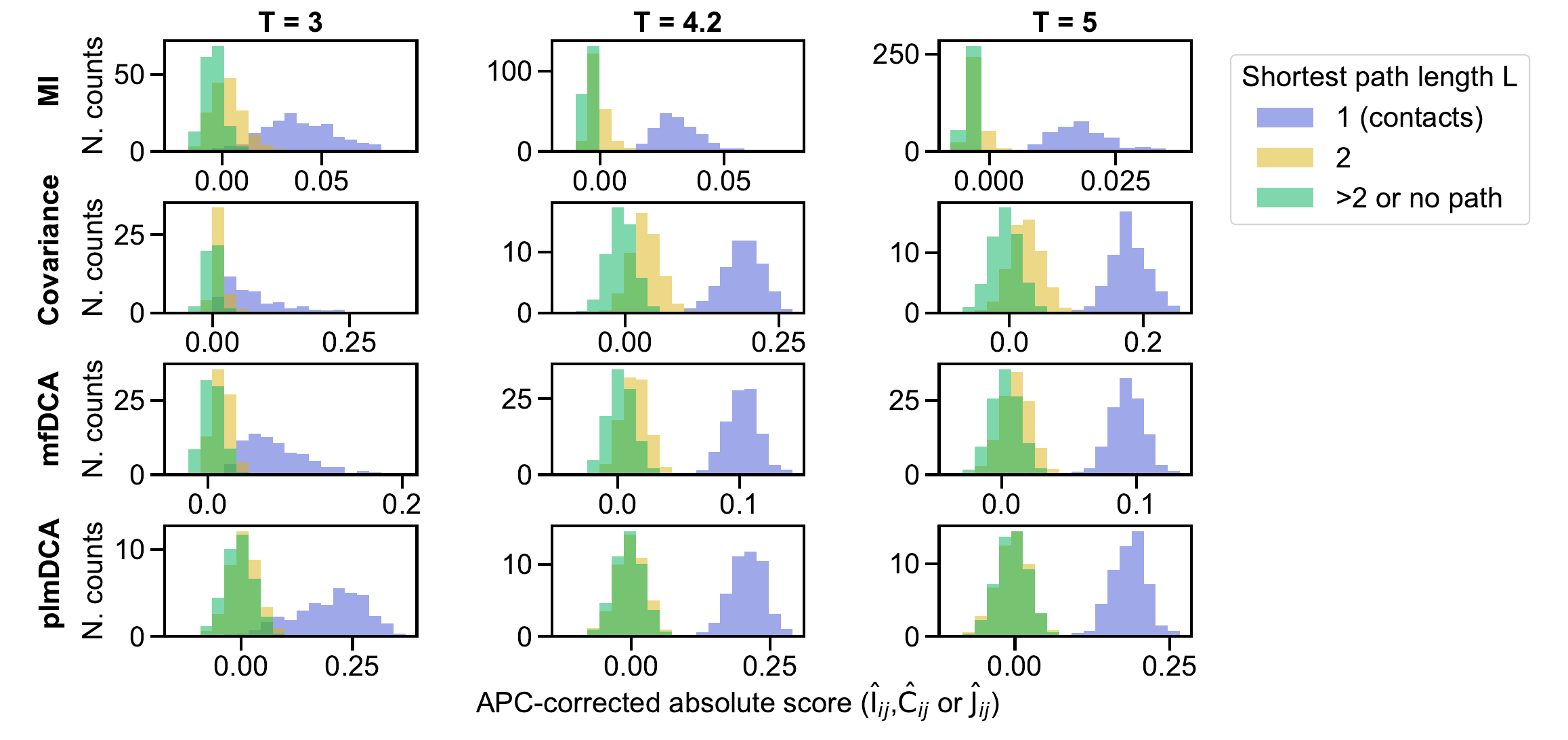}
\caption{\textbf{Impact of graph properties on APC-corrected absolute scores for equilibrium sequences.} Same as in Figure~\ref{fig:HistogramsSPL} except that the APC-corrected absolute values of the scores are reported instead of the raw scores.}
\label{fig:HistogramsSPL_APC}
\end{figure}

\begin{figure}[h!]
\centering
\includegraphics[scale = 0.35]{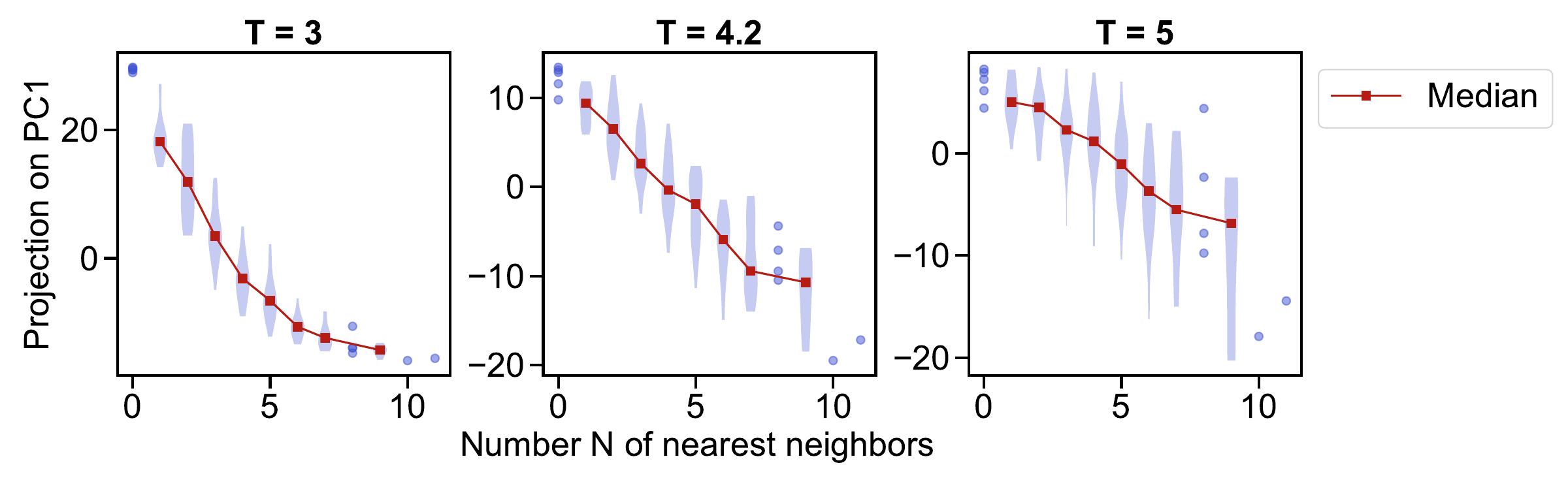}
\caption{\textbf{Impact of graph properties on data covariance for equilibrium sequences.} Violin plots of the projection of each site on the first principal component are shown versus the number $N$ of nearest neighbours of these sites. In each panel, principal component analysis (PCA) is performed on a data set of equilibrium sequences (where sequences are features and sites are observations -- in other words, the matrix of covariances between sequences, which is an $M\times M$ size matrix, where $M$ is the number of sequences, is diagonalised). The first principal component is the direction of largest variance of the sites (top ``eigensite''). Three data sets of 2048 sequences each are sampled at equilibrium at three temperatures: $T=3<T_C$ (left), $T=4.2\simeq T_C$ (center), and $T=5>T_C$ (right), employing the cluster algorithm (see Figure \ref{fig:Magnetisation_vs_tau}), and using the same contact map (Erd\H{o}s-Rényi graph) as in Figure~\ref{fig:tpfrac_vs_mu_sparse}. Ensembles of 5 data points or fewer are represented with individual markers instead of Gaussian kernel-smoothed histograms.}
\label{fig:PCA_NN}
\end{figure}

\clearpage

\section{Impact of phylogeny on coevolution scores and on conservation}

\begin{figure}[h!]
\centering
\includegraphics[scale = 0.38]{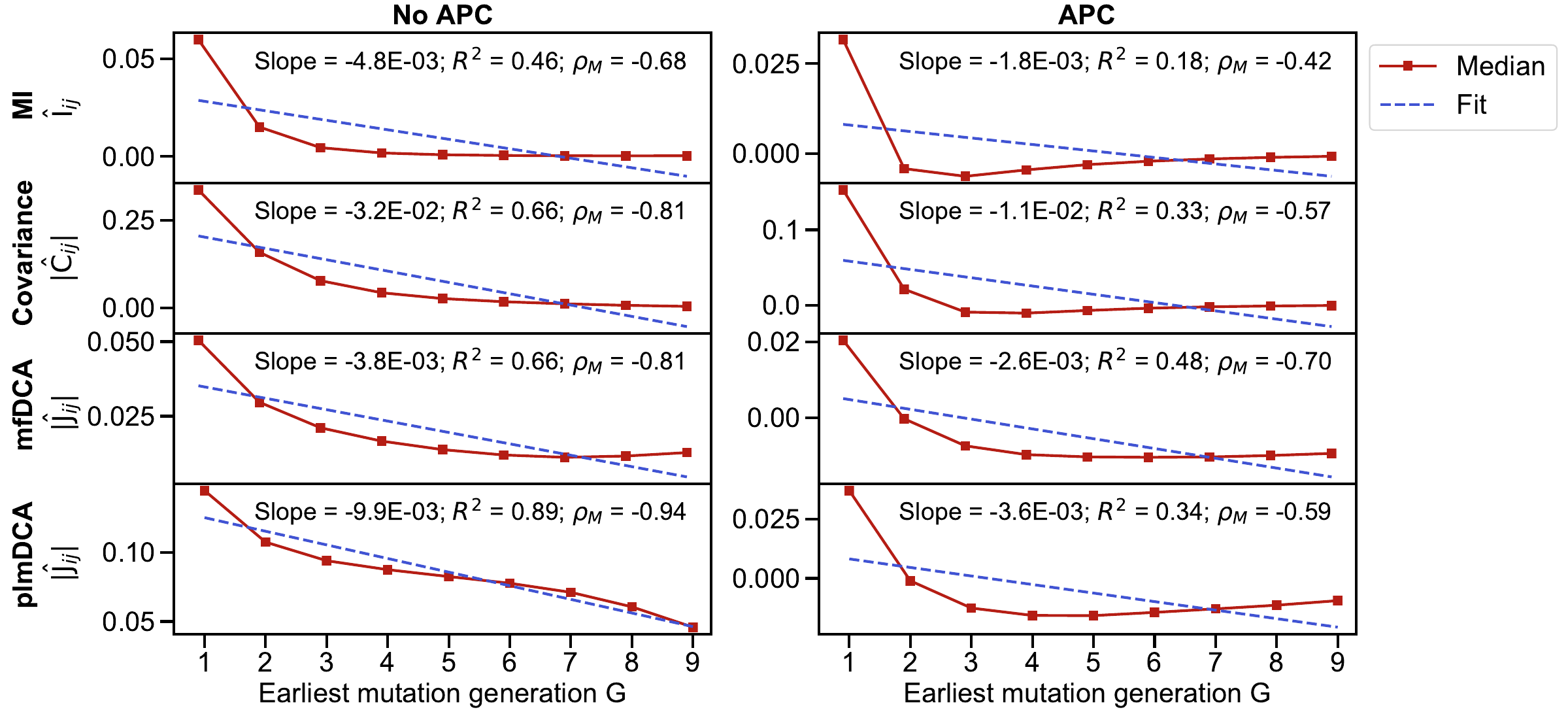}
\caption{\textbf{Impact of phylogeny on the median of the coevolution scores.} The median of the violin plots in Figure \ref{fig:Allcouplings_vs_emg} is plotted versus the earliest generation $G$ without (left panel) and with (right panel) the APC correction on the scores. A linear fit of the median is performed in each case, and the slope of the fit, the coefficient of determination $R^2$  and the Pearson correlation $\rho_M$ between the median score and $G$ are shown. The data is the same as in Figure \ref{fig:Allcouplings_vs_emg}.}
\label{fig:median_vs_emg_fit}
\end{figure}

\begin{figure}[h!]
\centering
\includegraphics[scale = 0.32]{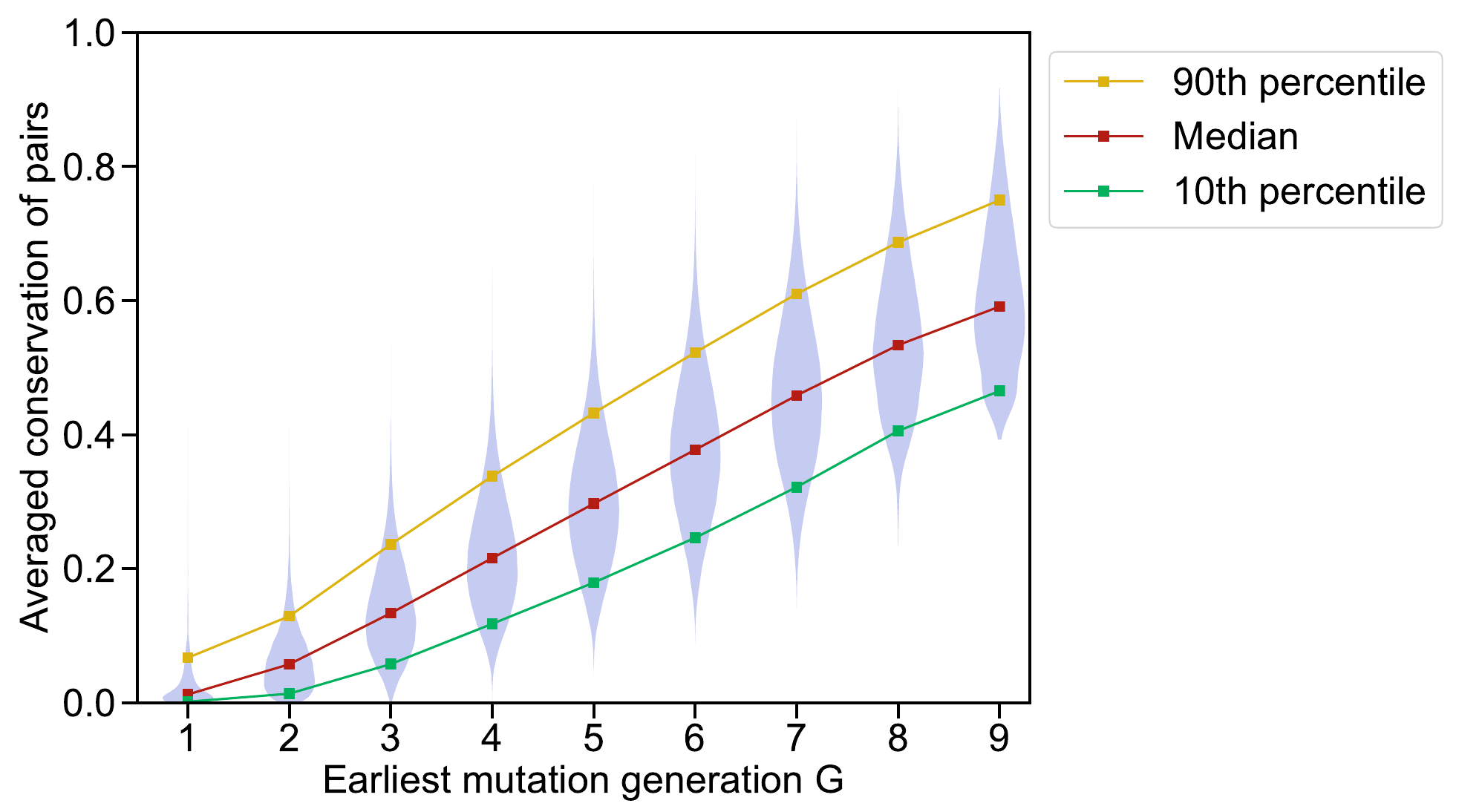}
\caption{\textbf{Impact of phylogeny on conservation scores.} Violin plots of the average conservation scores of all pairs of sites $(i,j)$ are shown versus the earliest generation $G$ where both $i$ and $j$ have mutated with respect to the ancestral sequence. The conservation of $i$ is defined as the difference of the maximum possible entropy of a site (1 bit) and the estimated entropy of $i$ (in bits), computed using frequencies instead of probabilities. The average conservation of a pair $(i,j)$ is the mean of the scores of $i$ and $j$. Data is generated as in Figure~\ref{fig:tpfrac_vs_mu_sparse}, using the same contact map (Erd\H{o}s-Rényi graph), and parameters $T=5$ and $\mu=5$. Couplings are inferred on 100 data sets comprising 2048 sequences each, and aggregated. The squared Pearson correlation between average conservation and earliest mutation generation is $r^2=0.53$.}
\label{fig:Averaged_conservation_vs_emg}
\end{figure}

\newpage

\clearpage
\section{Natural and more realistic data}

\begin{figure}[htbp]
\centering
%\includesvg[scale = 0.15]{ContactMaps_4proteinfamilies_Nat_EqbmDCA_PhylobmDCA_%MI_phyloweights_APC.svg}
\includegraphics[scale = 0.15]{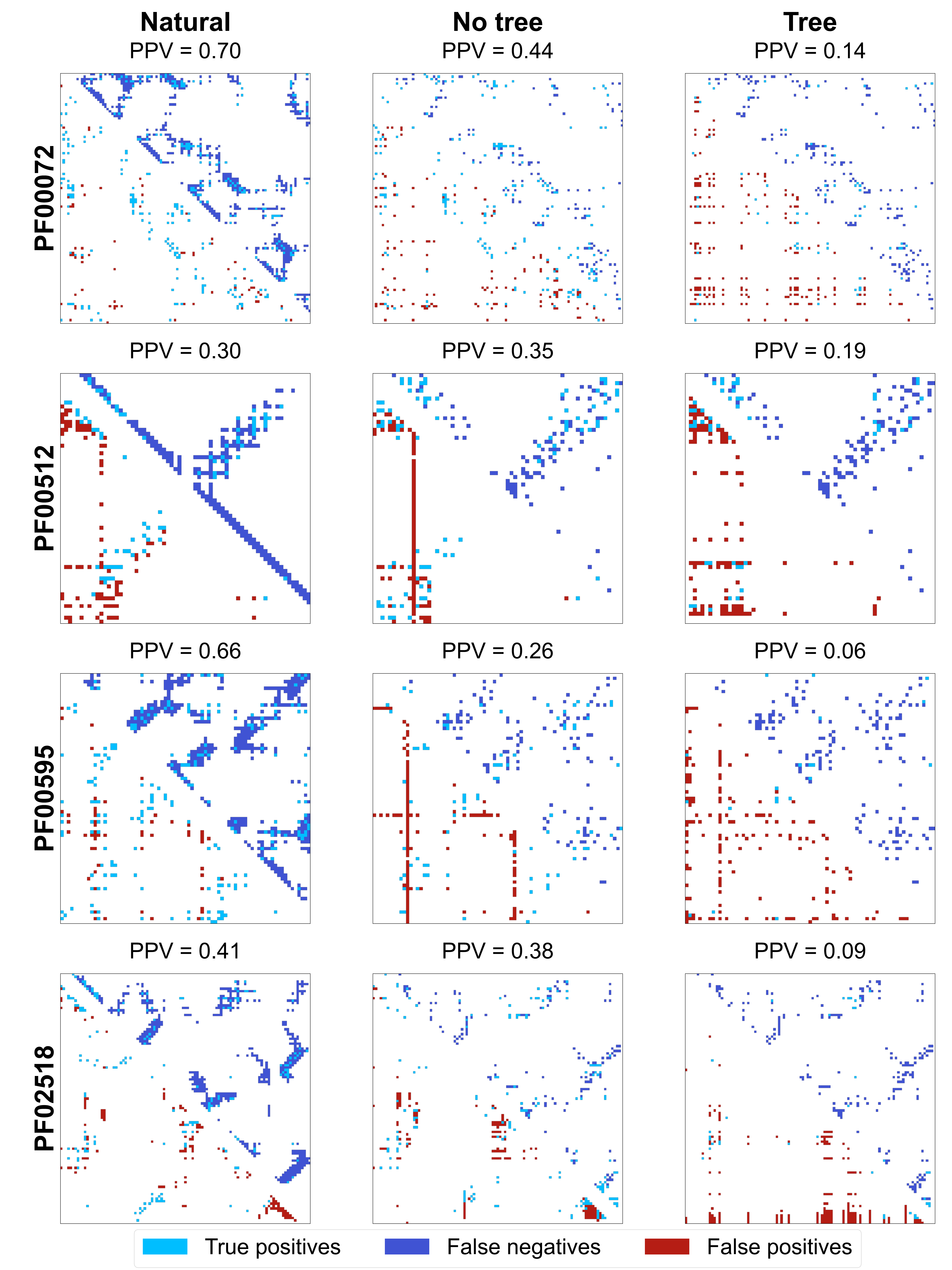}
\caption{\textbf{Contact maps predicted by mutual information from natural and realistic MSAs.}
Each panel shows the ground truth in the upper triangular part and the mutual information (MI)-based inference in the lower triangular part. Inference is performed using the phylogenetic reweighting (PR) and APC corrections. For each Pfam family, the left column shows an experimental contact map as ground truth and the MI-based inference performed on the natural MSA. Experimental contact maps use the PDB structures listed in table \ref{tab:MSA}, with an all-atom Euclidean distance cutoff of $\SI{8}{\angstrom}$, excluding residue pairs at positions $i,j$ with $|i-j| \le 4$. The middle (resp.\ right) column presents the contact maps inferred from the natural sequences by bmDCA as ground truth and the MI-based inference performed on synthetic sequences generated at equilibrium (resp.\ along a phylogenetic tree inferred from the natural MSA) using the bmDCA model inferred from the natural MSA (see Methods, ``Generating more realistic sequences''). In all cases, the top 2$\ell$ scores are predicted as contacts, and the associated PPV (or true positive fraction) is shown.}
\label{fig:contactmaps_realisticdata_miapc}
\end{figure}

\begin{figure}[htbp]
\centering
%\includesvg[scale = 0.15]{ContactMaps_4proteinfamilies_Nat_EqbmDCA_PhylobmDCA_%plmDCA_phyloweights_APC.svg}
\includegraphics[scale = 0.15]{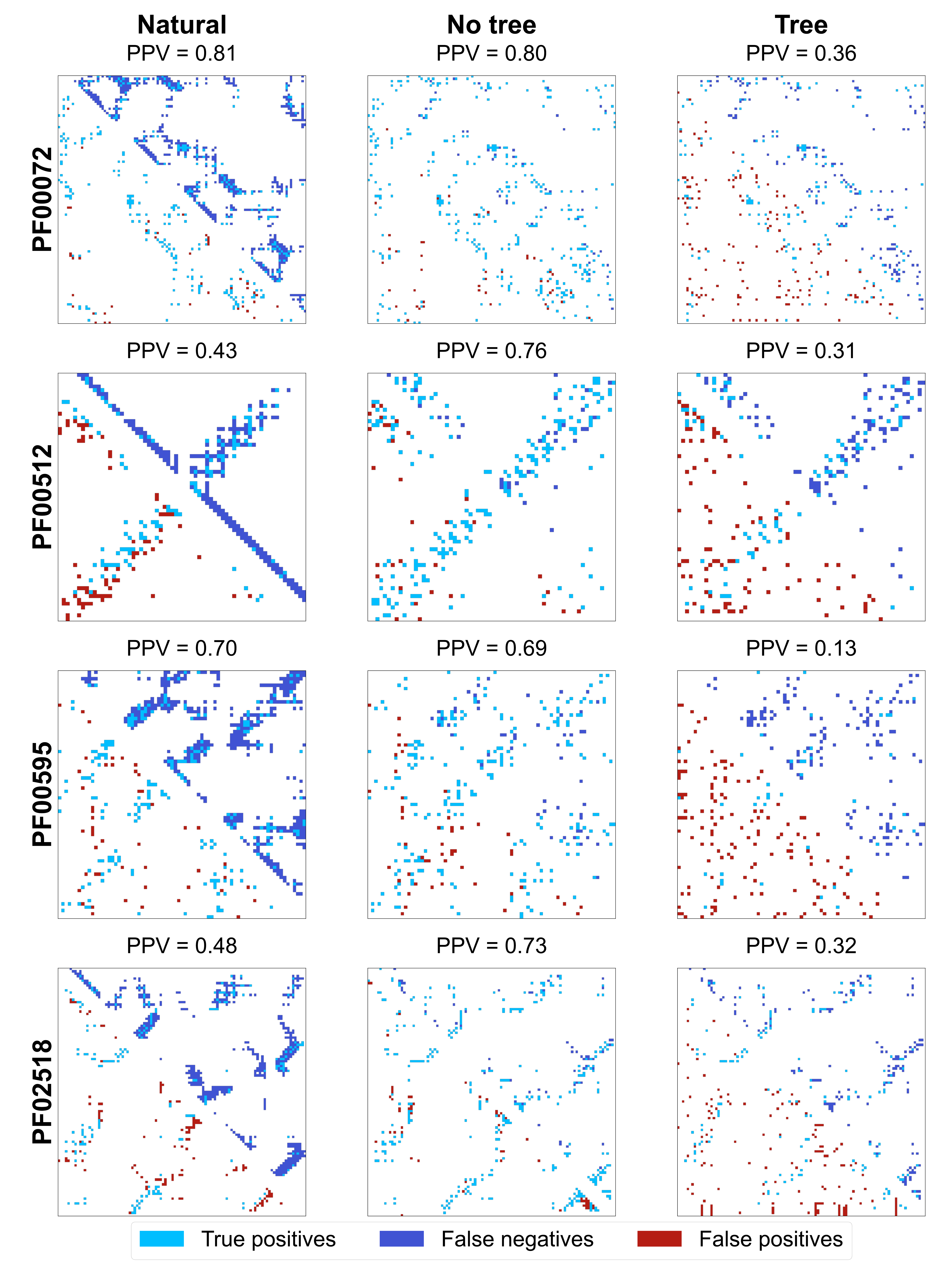}
\caption{\textbf{Contact maps predicted by plmDCA from natural and more realistic MSAs.}
Same as in figure~\ref{fig:contactmaps_realisticdata_miapc}, but using plmDCA instead of MI to infer contacts. Here too, phylogenetic reweightings are used, as well as the APC correction.}
\label{fig:contactmaps_realisticdata_plm}
\end{figure}

\begin{table}[h!]
   \centering
   \begin{tabular}{@{}llclllcllcll@{}}
       \toprule
       &  && \multicolumn{3}{c}{MSA} & &  \multicolumn{2}{c}{PDB structure}&& \multicolumn{2}{c}{Contact density} \\
       \cmidrule{4-6} \cmidrule{8-9} \cmidrule{11-12}
       \multicolumn{1}{c}{Pfam ID} & \multicolumn{1}{c}{Family name}&& \multicolumn{1}{c}{$\ell$} & \multicolumn{1}{c}{$M$} & \multicolumn{1}{c}{$M_\mathrm{eff}^{(PR)}$} && \multicolumn{1}{c}{ID} & \multicolumn{1}{c}{Resol.}&&  $\SI{4}{\angstrom}$ & $\SI{8}{\angstrom}$\\
       \midrule

       PF00072 & Response\_reg  && 112 & 73063 &15090 && 3ILH & $\SI{2.59}{\angstrom}$ &&0.02 & 0.13 \\
       PF00512 & HisKA  && 66 & 154998 &8980 && 3DGE & $\SI{2.80}{\angstrom}$ && 0.01 & 0.13\\
       PF00595 & PDZ  && 82 & 71303 &1419 && 1BE9 & $\SI{1.82}{\angstrom}$&&0.04 & 0.18 \\

       PF02518 & HATPase\_c && 111 & 80714 &16058 && 3G7E & $\SI{2.20}{\angstrom}$ && 0.02 & 0.11\\

       \bottomrule
   \end{tabular}
\caption{\textbf{Pfam families and MSAs considered in this work.} For each family, we considered the Pfam full alignments (``MSA''). For each of these MSAs, we report the length $\ell$ (number of amino acid sites), depth $M$ (number of sequences) and the effective depth $M^{(PR)}_{\mathrm{eff}}$ from the phylogenetic reweighting done in the \texttt{PlmDCA} package (\href{url}{https://github.com/pagnani/PlmDCA}). The PDB structures used as reference structures for each Pfam family, and their resolutions, are reported.  The density of the experimental contact maps (i.e.\ fraction of residue pairs $i,j$ that are in contact) is also shown, with two all-atom Euclidean distance cutoffs at $\SI{4}{\angstrom}$ and $\SI{8}{\angstrom}$, excluding residue pairs at positions $i,j$ with $|i-j| \le 4$.   \label{tab:MSA}}
\end{table}

\begin{table}[h!]
   \centering
   \begin{tabular}{@{}cccccccccc@{}}
       \toprule
        & \multicolumn{4}{c}{MI} & & \multicolumn{4}{c}{plmDCA}   \\
       \cmidrule{2-5} \cmidrule{7-10}
       \multicolumn{1}{c}{Pfam ID} & \multicolumn{1}{c}{Raw} & \multicolumn{1}{c}{PR} &\multicolumn{1}{c}{APC}&\multicolumn{1}{c}{PR+APC}&&\multicolumn{1}{c}{Raw} & \multicolumn{1}{c}{PR} & \multicolumn{1}{c}{APC} &\multicolumn{1}{c}{PR+APC}  \\
       \midrule
       PF00072 &0.36&0.52&0.59& 0.70&  &0.62&0.74&0.77&0.81\\
       PF00512 &0.21&0.27&0.33&0.30& &0.33&0.38&0.42&0.43\\
       PF00595 &0.22&0.46&0.53&0.66& &0.27&0.39&0.52&0.70\\
       PF02518 &0.22&0.25&0.39&0.41& &0.34&0.42&0.48&0.48\\
       \bottomrule
   \end{tabular}
\caption{\textbf{Impact of phylogenetic corrections on contact prediction in natural MSAs.} The four Pfam protein families introduced in Table~\ref{tab:MSA} were used. Contacts are predicted using mutual information (local) and plmDCA (global), without or with two different phylogenetic corrections. Performance of contact prediction is assessed as the fraction of true positive contacts among $N_\textrm{pred}=2\ell$ predicted contacts (see Table~\ref{tab:ppv_fp_nat_realistic_data}), which coincides with the PPV. ``Raw" means that no phylogenetic correction is used; ``PR" indicates that the phylogenetic reweighting correction of the frequencies is used; ``APC" means that the APC correction is used; ``PR+APC" means that both phylogenetic reweighting and APC are used, which is standard in DCA.
\label{tab:ppv_natseq}}
\end{table}

\begin{table}[h!]
   \centering
   \begin{tabular}{@{}cccccccccccc@{}}
       \toprule
        & \multicolumn{2}{c}{Method}&& \multicolumn{2}{c}{Natural}& & \multicolumn{2}{c}{Equilibrium}& &\multicolumn{2}{c}{Tree}  \\ \cmidrule{2-3}
       \cmidrule{5-6} \cmidrule{8-9}\cmidrule{11-12}
       \multicolumn{1}{c}{Pfam ID}&Score&Correction&\multicolumn{1}{c}{$N_\textrm{pred}$} & \multicolumn{1}{c}{PPV} & \multicolumn{1}{c}{$\mathrm{FP}_{L=2}$}& &\multicolumn{1}{c}{PPV}&\multicolumn{1}{c}{$\mathrm{FP}_{L=2}$}& &\multicolumn{1}{c}{PPV} & \multicolumn{1}{c}{$\mathrm{FP}_{L=2}$}  \\
       \midrule
       \multirow{4}{*}{PF00072}& MI&Raw& 224&0.36&89& &0.23&48& &0.13&32 \\
                       & MI &PR+APC& 224&0.70&60& &0.44&34& &0.14&30 \\
                & plmDCA&Raw& 224&0.62&55& &0.65&33& &0.26&21 \\
                & plmDCA &PR+APC& 224&0.81&30& &0.80&19& &0.36&22 \\
       
        \midrule
            \multirow{4}{*}{PF00512}& MI&Raw& 132&0.21&49& &0.32&36& &0.14&24 \\
                & MI&PR+APC& 132&0.30&50& &0.35&20& &0.19&40 \\
                & plmDCA&Raw& 132&0.33&47& &0.55&29& &0.18&29 \\
                & plmDCA&PR+APC& 132&0.43&46& &0.76&12& &0.31&27 \\
       \midrule
       \multirow{4}{*}{PF00595}& MI&Raw& 164&0.22&58& &0.41&54& &0.06&20 \\
                & MI &PR+APC& 164&0.66&51& &0.26&14& &0.06&15 \\
                & plmDCA &Raw& 164&0.27&55& &0.62&34& &0.07&33 \\
                & plmDCA &PR+APC& 164&0.70&38& &0.69&17& &0.13&20 \\
       \midrule
       \multirow{4}{*}{PF02518}& MI &Raw& 222&0.22&73& &0.22&36& &0.12&19 \\
                & MI&PR+APC& 222&0.41&66& &0.38&47& &0.09&11 \\
                & plmDCA&Raw& 222&0.34&71& &0.58&32& &0.27&26 \\
                & plmDCA &PR+APC& 222&0.48&61& &0.73&18& &0.32&13 \\
       \bottomrule
   \end{tabular}
\caption{\textbf{Contact prediction performance and indirect correlations in natural and realistic data sets.} Performance of contact prediction is assessed as the fraction of true positive contacts among $N_\textrm{pred}=2\ell$ predicted contacts, which coincides with the PPV. Mutual information (local) and plmDCA (global) are used to predict contacts, either with no phylogenetic correction (``Raw") or with both phylogenetic reweighting and APC (``PR+APC"). For each of the four Pfam families on Table~\ref{tab:MSA}, three different data sets are compared: natural MSAs, equilibrium bmDCA-generated MSAs and MSAs generated using bmDCA along an inferred phylogenetic tree (see Methods, ``Generating more realistic sequences''). In addition to the PPVs, the number of false positive pairs with a shortest path length of $L=2$ (i.e., indirect correlations of order 1) are shown. 
\label{tab:ppv_fp_nat_realistic_data}}
\end{table}

\end{document}